\documentclass[english, aps, prb, twocolumn, superscriptaddress, secnumarabic, showkeys, nofootinbib, floatfix]{revtex4}

\usepackage[english]{babel}

\usepackage{amssymb, amsmath, bm, soul}

\usepackage[pdftex]{graphicx}
\usepackage[caption=false]{subfig}
\usepackage{color}
\usepackage[usenames, dvipsnames]{xcolor}
\usepackage[linktoc=page, colorlinks=true, linkcolor=MidnightBlue, citecolor=OliveGreen, urlcolor=OliveGreen]{hyperref} 

\newcommand{\x}{\mathbf{r}}
\newcommand{\Jc}{J_\mathrm{c}}
\newcommand{\Ic}{I_\mathrm{c}}
\newcommand{\Jdp}{J_\mathrm{dp}}
\newcommand{\Jext}{J}
\newcommand{\Tc}{T_\mathrm{c}}
\newcommand{\Tcb}{T_\mathrm{c,bulk}}
\newcommand{\Ec}{E_\mathrm{c}}
\newcommand{\Hct}{H_\mathrm{c2}}
\newcommand{\p}{\mathbf{p}}
\newcommand{\fbulk}{f}
\newcommand{\fin}{f_\mathrm{in}}
\newcommand{\fout}{f_\mathrm{out}}
\newcommand{\Lin}{l_\mathrm{in}}
\newcommand{\Lout}{l_\mathrm{out}}
\newcommand{\opt}{\mathrm{opt}}
\newcommand{\fL}{F_\mathrm{\scriptscriptstyle L}}

\newcommand{\subfignum}[1]{{\color{MidnightBlue}#1}}

\begin{document}

\title{Edge effect pinning in mesoscopic superconducting strips with non-uniform distribution of defects}

\author{Gregory J. Kimmel}
\affiliation{Materials Science Division, Argonne National Laboratory, 9700 S Cass Av, Lemont, IL 60439, USA}
\affiliation{Department of Engineering Sciences and Applied Mathematics, Northwestern University, 633 Clark St, Evanston, IL 60208, USA}

\author{Andreas Glatz}
\affiliation{Materials Science Division, Argonne National Laboratory, 9700 S Cass Av, Lemont, IL 60439, USA}
\affiliation{Department of Physics, Northern Illinois University, DeKalb, IL 60115, USA}

\author{Valerii M. Vinokur}
\affiliation{Materials Science Division, Argonne National Laboratory, 9700 S Cass Av, Lemont, IL 60439, USA}

\author{Ivan A. Sadovskyy} 
\affiliation{Materials Science Division, Argonne National Laboratory, 9700 S Cass Av, Lemont, IL 60439, USA}
\affiliation{Computation Institute, University of Chicago, 5735 S Ellis Av, Chicago, IL 60637, USA}

\date{November 1, 2018}

\begin{abstract}
Transport characteristics of nano-sized superconducting strips and bridges are determined by an intricate interplay of surface and bulk pinning. In the limiting case of a very narrow  bridge, the critical current is mostly defined by its surface barrier, while in the opposite case of very wide strips it is dominated by its bulk pinning properties. Here we present a detailed study of the intermediate regime, where the critical current is determined, both, by randomly placed pinning centers and by the Bean-Livingston barrier at the edge of the superconducting strip in an external magnetic field. We use the time-dependent Ginzburg-Landau equations to describe the vortex dynamics and current distribution in the critical regime. Our studies reveal that while the bulk defects arrest vortex motion away from the edges, defects in their close vicinity promote vortex penetration, thus suppressing the critical current. We determine the spatial distribution of the defects optimizing the critical current and find that it is in general non-uniform and asymmetric: the barrier at the vortex-exit edge influence the critical current much stronger than the vortex-entrance edge. Furthermore, this optimized defect distribution has a more than 30\% higher critical current density than a homogeneously disorder superconducting film.
\end{abstract}

\keywords{
	Type-II superconductivity,
	critical current,
	vortex trapping,
	Bean-Livingston barrier,
	time-dependent Ginzburg-Landau model
}

\maketitle


\section{Introduction}

Immobilizing magnetic vortices and thus preventing dissipation under applied currents is one of the major objectives for realizing applications of {type-II} superconductivity.\cite{Blatter:1994, Holesinger:2008, Malozemoff:2012, Kwok:2016} Typically, this vortex pinning is achieved by introducing structural inhomogeneities in the bulk of the material. Recently, it has been recognized that geometric pinning utilizing surface and geometrical barriers for controlling the entrance or exit of vortices in and out of mesoscopic superconductors and superconducting strips can be extremely efficient.\cite{Stan:2004, Zeldov:1994, Kuit:2008, Vodolazov:2013, Willa:2014, Papari:2016, Wang:2017, Sadovskyy:2018} Appreciable enhancement of superconducting parameters in strips was recently observed experimentally and explained in terms of surface (edge) superconductivity.\cite{Cordoba:2013, Berdiyorov:2012} One could conclude from these experiments that surfaces may provide one of the most important pinning mechanisms in strips and mesoscopic systems.\cite{Kupriyanov:1974, Tahara:1990, Benk:1998} At the same time, it was observed that the introduction of point-like or cylindrical defects near the surface can be detrimental to the effectiveness of surface barriers\cite{Bean:1962, Bean:1964} since they promote easier vortex penetration across the surface.\cite{Schuster:1994} Hence the effect of structural disorder is two-fold: it arrests the vortex dynamics in the bulk, but `contaminates' surface pinning.\cite{Ivanchenko:1983, Koshelev:2001, Gregory:2001, Bush:2002} Both effects are important in an intermediate width regime where each mechanism contributes to the critical current, which is the largest possible applied current at which magnetic vortices are immobile.

In the case of narrow strips with widths on the order of the superconducting coherence length, the critical current is mostly defined by its surface barrier and phase slips across the strip are important,\cite{Ovchinnikov:2015, Kimmel:2017b} while for very wide strips, the critical current is dominated by its bulk pinning properties. This sets the quest for optimizing artificially manufactured disorder in geometrically restricted systems to take advantage of a potentially constructive interplay of bulk and surface pinning mechanisms.

\begin{figure*}
	\vspace{-2mm}
	\subfloat{\includegraphics[width=150mm]{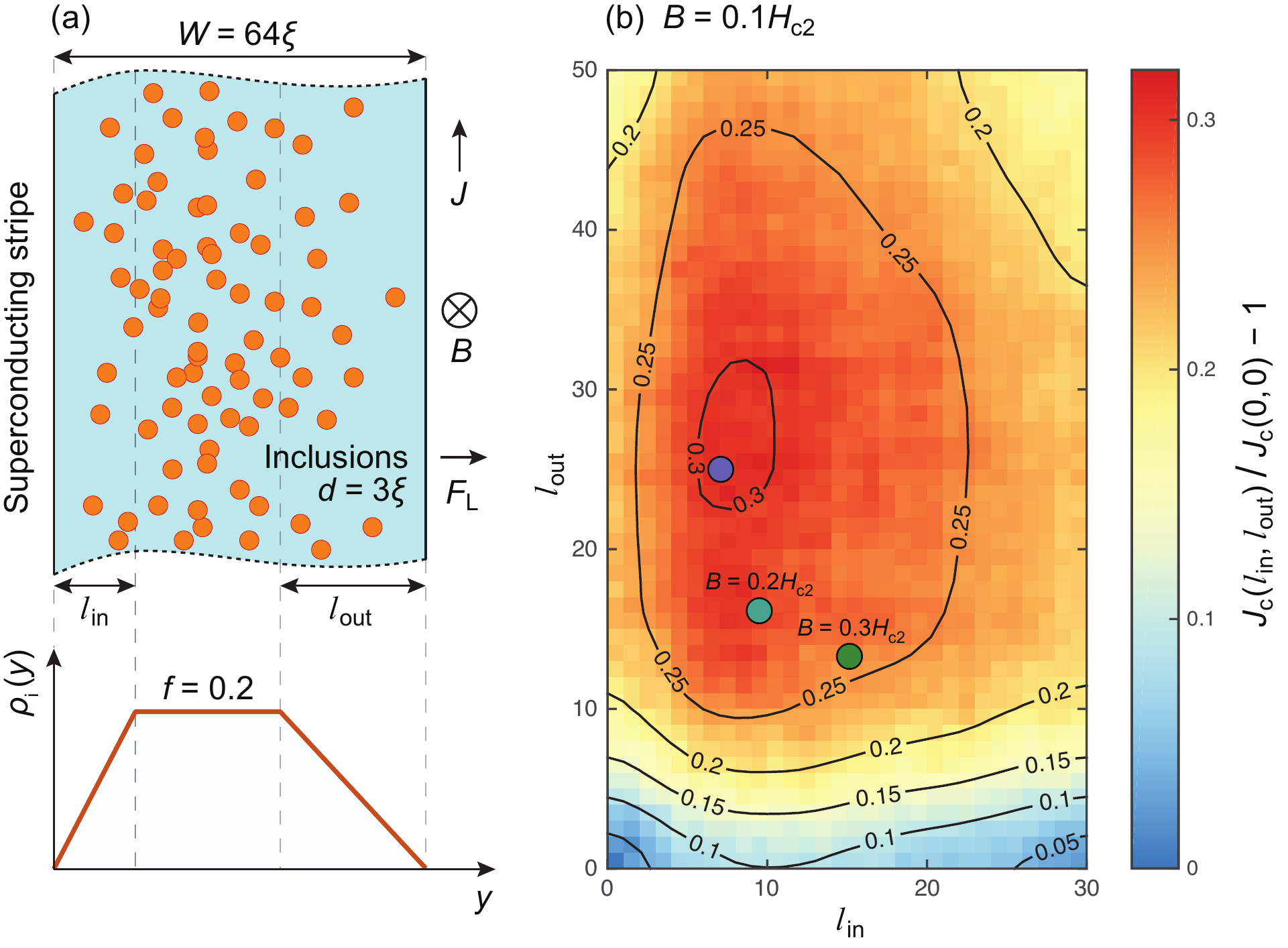}\label{fig:setup}}
	\subfloat{\label{fig:Jc_in_out}}
	\vspace{-2mm}
	\caption{
		\subfignum{(a)}~Two-dimensional superconducting strip of width $W = 64\xi$ 
		with non-homogeneous inclusion distribution. The current~$\Jext$ is applied 
		vertically (along the $x$-axis), the magnetic field $B$ is perpendicular 
		to the figure plane, and the resulting Lorentz force~$\fL$ acts to the right 
		(along the $y$-axis). The sample has a length of $L = 1024\xi$ 
		with quasi-periodic boundary conditions in the $x$ direction; in the $y$ 
		direction, we have open boundary conditions, i.e., superconductor-vacuum 
		surfaces. The strip contains (uncorrelated) randomly placed circular inclusions 
		of diameter $d = 3\xi$. The density of these inclusions depends on~$y$: 
		in the middle of the sample, the volume fraction occupied by inclusions 
		is $f = 0.2$, which corresponds approximately to conditions for the maximum 
		possible critical current density in bulk samples. The density of the inclusion 
		$\rho_\mathrm{i}(y)$ decreases linearly near the sample boundaries 
		(see bottom plot): within a region of width $\Lin$ at the boundary where 
		vortices enter the sample and $\Lout$ at the boundary where vortices 
		leave the sample.
		\subfignum{(b)}~The critical current $\Jc$ as a function of $\Lin$ and 
		$\Lout$ normalized by $\Jc(0,0)$ at applied magnetic field $B = 0.1\Hct$. 
		The critical current is increased by $\sim 30\%$ for finite $\Lin$ and $\Lout$ 
		compared to the critical current from a homogeneous defect distribution 
		($\Lin = \Lout = 0$). The values of $\Lin$ and $\Lout$ corresponding 
		to the maximum of the critical current $\Jc(\Lin,\Lout)$ are shown 
		by colored circles for $B = 0.1\Hct$, $0.2\Hct$, and $0.3\Hct$. 
		The effect is asymmetric and depends on the direction of vortex motion. 
		The maximum is indicated by a (blue) circle. Corresponding maxima 
		for fields  $0.2\Hct$ and  $0.3\Hct$ are indicated by (cyan and green) circles, 
		marked by the field value.
	}
	\label{fig:setup_Jc_in_out}
\end{figure*}

The present article addresses this problem. To this end, we design an approach allowing us to optimize the concentration and spatial distribution of the bulk point defects in order to achieve the maximum possible critical current taking into account the interplay between the surface barrier blocking penetration of vortices into a superconductor and bulk defects arresting the vortex motion in the interior of the sample. We consider experimentally important systems: superconducting wires having the shape of tapes with widths on the order of a few tens of the superconducting coherence length.\cite{Malozemoff:2012} In order to calculate the critical current for a given arrangement of pins (pinscape), we use a solver for the time-dependent Ginzburg-Landau (TDGL) equation for {type-II} superconductors.\cite{Sadovskyy:2015a} This approach describes the vortex dynamics sufficiently well in superconductors near the vicinity of the critical temperature and is capable of reproducing experimental critical currents for a given pinscape.\cite{Berdiyorov:2006, Sadovskyy:2016a, Sadovskyy:2016b, Sadovskyy:2017}

The article is organized as follows. We introduce Ginzburg-Landau model and review known cases in Sec.~\ref{sec:model}. We present the results for the strip containing non-uniformly distributed defects in Sec.~\ref{sec:results}. We discuss and summarize our results in Sec.~\ref{sec:conclusions}.

\section{Model} \label{sec:model}

We consider a two-dimensional superconducting strip, infinite in the $x$ direction and a finite width $W$, which is appreciably larger than the superconducting coherence length, $\xi$, but less than the London penetration depth, $\lambda$. The edges at $y = 0$ and $y = W$ set the positions of the surface barriers. Bulk defects are introduced by spatial modulation of the transition temperature, $\Tc(\x)$. To evaluate the critical current for the system, we use the TDGL equation, which simulates the dynamic behavior of the complex superconducting order parameter $\psi = \psi(\x, t)$:
\begin{equation} 
	\bigl( \partial_t + i\mu \bigr) \psi 
	= \epsilon(\x) \psi - |\psi |^2\psi + 
	\bigl( \nabla - i\mathbf{A} \bigr)^2\psi + \zeta(\x, t).
	\label{eq:GL} 
\end{equation} 
Here $\mu = \mu(\x, t)$ is the scalar potential, $\bf A$ is the vector potential generating the external magnetic field $\mathbf{B} = \nabla \times \mathbf{A}$, and $\zeta(\x, t)$ is a temperature-dependent $\delta$-correlated Langevin thermal noise term. The unit of length is defined by the superconducting coherence length~$\xi = \xi(T)$ at a given temperature $T$ and the unit of the magnetic field is the upper critical field $\Hct = \Hct(T)$. Defects in the bulk are realized through the parameter $\epsilon(\x) = [\Tc(\x) - T] / [\Tcb - T]$, where $\Tcb$ is the transition temperature for the clean sample. We solve the TDGL equation in the infinite-$\lambda$ limit, allowing us to use the gauge $\mathbf{A} = (-B_z$, $0$, $0) y$ for the vector potential.

We solve Eq.~\eqref{eq:GL} numerically by discretizing the system on a regular grid with mesh size of half a coherence length and integration of time using an implicit massively parallel iterative solver, see Ref.~\onlinecite{Sadovskyy:2015a} for implementation details. We consider the model system shown in Fig.~\subref{fig:setup}, where the two-dimensional superconducting strip lies in the $xy$ plane with quasi-periodic boundary conditions imposed in $x$ direction and open boundary conditions in $y$ direction (i.e., the $y$ component of the current has to obey $J_y = 0$ at these boundaries corresponding to a superconductor-vacuum surface). The magnetic field $B$ is applied in $z$ direction and the external current $\Jext$ is applied in the $x$ direction. In this case, the Lorentz force drives vortices in $+y$ direction (i.e., vortices enter the domain from $y = 0$ and exit at $y = W$).

The current density,
\begin{equation}
	\mathbf{J} 
	= \frac{3\sqrt{3}}{2} \Bigl\{ 
		\mathrm{Im} \bigl[ \psi^*(\nabla - i\mathbf{A})\psi \bigr]
		- \nabla \mu 
	\Bigr\}
	\label{eq:J} 
\end{equation} 
is measured in units of the \textit{depairing current} $\Jdp = \Jdp(T)$. $\Jdp$ is the current at which the superconducting order parameter is suppressed to zero, or Cooper pairs are not stable anymore, i.e., superconductivity is completely destroyed.

The magnitude of the critical current in the presence of an external magnetic field is controlled by inclusion patterns, which are small non-superconducting islands immersed in the superconducting matrix. We tune the inclusion size (typically a few $\xi$) and their spatial distribution. 

To determine the magnitude of the critical current, we use a finite-electrical-field criterion. Specifically, we chose a certain small external electric field, $\Ec = 10^{-4} (3\sqrt{3}/2) \Jdp / \sigma$, where $\sigma$ is the normal conductivity, and adjust the external current, $\Jext$, to keep this electrical-field criterion on average. The time-averaged value of the external current in the steady state gives the critical current, $\Jc = \langle \Jext \rangle$. 

\begin{figure}
	\subfloat{\includegraphics[width=55mm]{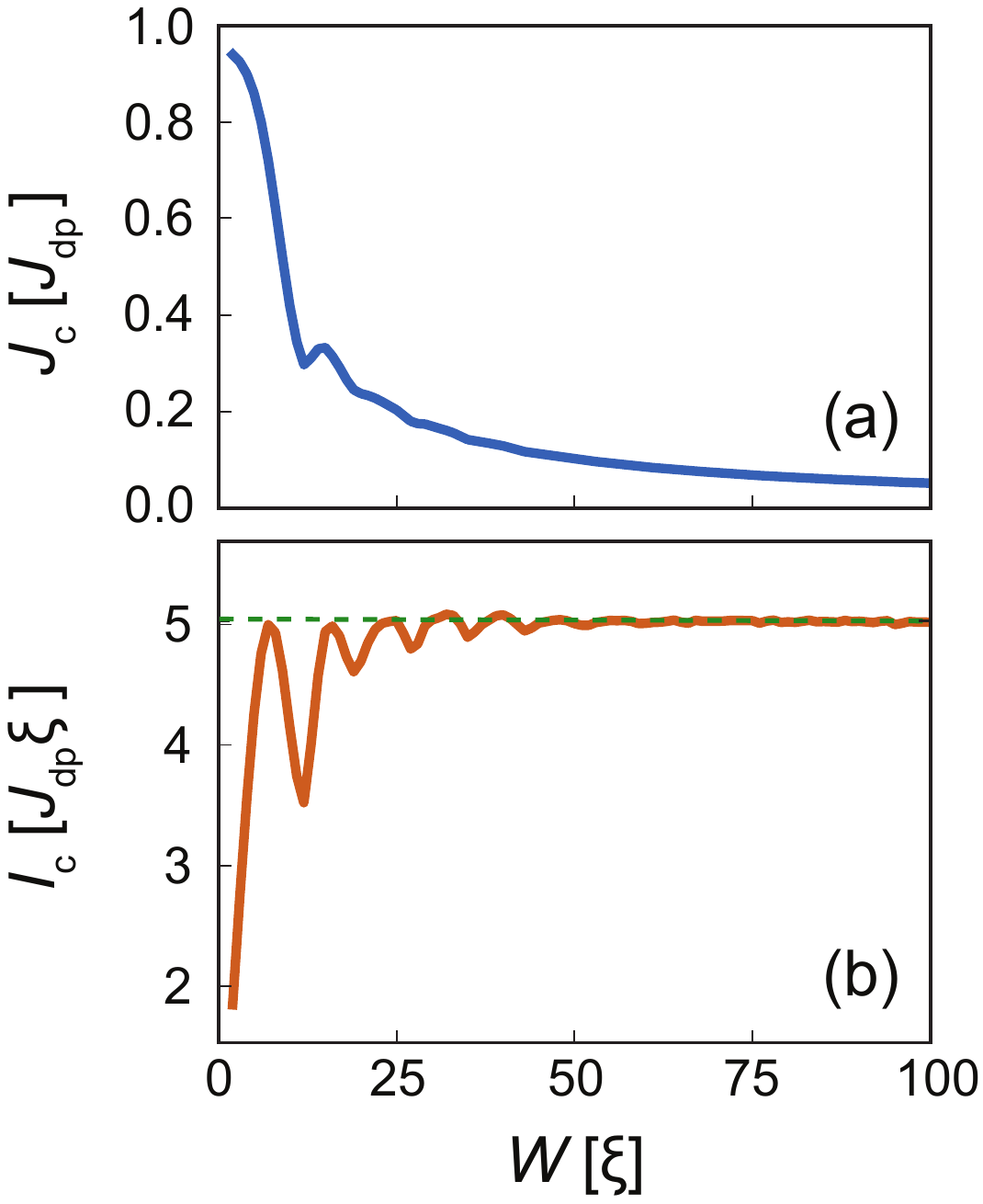}\label{fig:Jc_w}}
	\subfloat{\label{fig:Ic_w}}
	\vspace{-1mm}
	\caption{
		\subfignum{(a)}~Critical current density $\Jc$ and \subfignum{(b)}~critical 
		current $\Ic = \Jc W$ as a function of width $W$ of the ideal superconducting 
		strip containing no inclusions in magnetic field $B = 0.1\Hct$ applied 
		perpendicular to the strip. The critical current is defined by strip boundaries only 
		and saturates at $\Ic \approx 5\Jdp\xi$ (dashed line) for $W \gtrsim 64\xi$ 
		due to the absence of pinning potentials in the bulk. Certain vortex configurations 
		with few commensurate vortex rows (in particular the 4 and 5 row configuration) 
		are very stable due to geometrical pinning and can have larger critical currents 
		than the saturation value, see Ref.~\onlinecite{Papari:2016}. Artifacts from 
		the constant voltage criterion, used to determine the critical current, for wide clean 
		strips are removed, see text.
	}
	\label{fig:Jc_Ic_w}
\end{figure}

\subsection{Clean strip}

We start with the two limiting situations: a clean strip and  bulk superconductor with defects.

The pinning force in a case of clean strip is defined by edges at $y = 0$ and $y = W$ with open (no-current) boundary conditions. These boundaries produce the Bean-Livingston barrier\cite{Bean:1962, Bean:1964, Brandt:1993, Burlachkov:1991, Burlachkov:1993, Burlachkov:1996} and arrange vortices in `rows' along the current direction.\cite{Papari:2016} The number of rows depends on the width of the strip $W$ and on the applied magnetic field $B$. At fixed magnetic field, the most stable configurations are achieved under commensurability conditions. Therefore upon changing the width, the number of the stable rows varies as well, leading to oscillations in the critical current density $\Jc(W)$, which are more pronounced in the total critical current $\Ic(W) = \Jc(W) W$ as shown in Figs.~\subref{fig:Jc_w} and \subref{fig:Ic_w}, respectively. The maxima are realized when the system can accommodate the number of vortices corresponding to the applied field and minima when the system is in between two stable vortex lattice configurations. These oscillations can be best observed for the first few vortex rows. For $W \gg 1$, the critical current $\Ic$ saturates at some certain value defined by the depinning forces of the two barriers and depends on the magnetic field. Note, that certain commensurate vortex configurations are very stable (in particular for 4 or 5 rows), such that the critical current for these configurations can be even larger than the saturation value. We remark that the method to determine the critical current described above is independent of $W$, which leads to small linear increase in the critical current with the width of the system as the critical current density saturates when the free-flow voltage (the free-flow regime is the regime of linear current-voltage  behavior where vortices are not pinned anymore)is equal to the chosen electric field cutoff (which determines the slope of increase). This artificial increase becomes recognizable for very wide systems and is therefore subtracted from the critical current in Fig.~\subref{fig:Ic_w}.

\subsection{Bulk superconductor}

In this case, the critical current associated with pinning vortices at non-superconducting defects depends on the defect properties (shape, size, concentration) and on the field strength (vortex density). In a  three-dimensional (3D) bulk {type-II} superconductor containing spherical particles and for a wide range of fixed applied magnetic fields, $0.02 \Hct < B < 0.2 \Hct$, the optimal critical current is achieved for particle diameters $d$ ranging from $2.5\xi$ to $4.5\xi$ and $15$--$20\%$ volume fraction occupied by particles.\cite{Koshelev:2016} For large inclusions of fixed diameter $d \geqslant 3\xi$, the field dependence of the critical current has shown peculiar peaks, associated with the inclusion's occupancy by multiple vortices.\cite{Willa:2018a, Willa:2018b} Similar results are observed in regular and random pinning configurations of circular (cylindrical) defects in two-dimensional (3D) systems.\cite{Sadovskyy:2017, Kimmel:2017a} Note that a 2D system with circular defects is comparable to a 3D system with columnar rather than spherical defects, see below.

\subsection{General case} Now, we consider geometrically confined 2D systems with circular defects. We design the pinning configuration within our model system with finite $W$ in the following way: (i)~the density of the non-superconducting columnar defects far away from the edges is the same as in the bulk case corresponding to the maximum possible critical current; (ii)~the density of non-superconducting defects near edges is linearly modulated towards the edges. We define the volume fraction $\rho_\mathrm{i}(y)$ occupied by defects of the same diameter $d$ as a function of $y$ which is given by
\begin{equation}
	\rho_\mathrm{i}(y) = \begin{cases} 
		\cfrac{y}{\Lin} \, f + \cfrac{\Lin \! - \! y}{\Lin} \, \fin, & y < \Lin, \vspace{0.6em} \\
		\fbulk, & \Lin \leqslant y \leqslant W \! - \Lout, \vspace{0.6em} \\
		\cfrac{W \! - \! y}{\Lout} \, \fbulk + \cfrac{\Lout \! + \! y \! - \! W}{\Lout} \, \fout, & y > W \! - \Lout.
	\end{cases}
	\label{eq:rho_general}
\end{equation}
In particular, the volume fraction of the defects changes linearly from $\fin$ to its bulk value $\fbulk$ at the distance $\Lin$ from the edge $y = 0$ where vortices enter the sample. On the opposite side of the sample $\rho_\mathrm{i}(y)$ changes from $\fbulk$ to $\fout$ at distance~$\Lout$.

\section{Results} \label{sec:results}

The surface barrier at the superconductor edges prevent vortices from entering and exiting the superconductor. As mentioned in the introduction, non-superconducting defects located at edges or in the vicinity of edges effectively reduce the Bean-Livingston barrier by creating weak spots for vortex penetration.\cite{Koshelev:2001} We study the interplay between the surface barrier and defect distribution profile $\rho_\mathrm{i}(y)$ by investigating the dependence of the critical current density, $\Jc$, on the parameters $\fbulk$, $\fin$, $\fout$, $\Lin$, $\Lout$, $d$, in a fixed magnetic field $B$  and fixed sample width $W \gg \Lin$, $\Lout$. 
Therefore, we start our numerical investigation with initial initial investigations of the full 6D optimization problem
\begin{equation}
	\p^\opt = \mathrm{arg} \max_\p \Jc(\p)
	\label{eq:opt}
\end{equation}
with control parameter set $\p = \{ \fbulk$, $\fin$, $\fout$, $\Lin$, $\Lout$, $d \}$ for different fixed magnetic fields using a particle swarm optimization routine.\cite{Kimmel:2017a} The resulting optimal parameter set $\p^\opt$ corresponds to the maximum critical current density $\Jc(\p^\opt)$.  These initial studies revealed that for the range of applied magnetic fields investigated in this paper, the optimal concentrations of the defects near the entrance and exit boundaries were zero, $\fin^\opt = \fout^\opt = 0$. This allows us to simplify the initial model density profile~\eqref{eq:rho_general} to
\begin{equation}
	\rho_\mathrm{i}(y) = \begin{cases} 
		\cfrac{y}{\Lin} \, \fbulk, & y < \Lin, \vspace{0.6em} \\
		\fbulk, & \Lin \leqslant y \leqslant W - \Lout, \vspace{0.6em} \\
		\cfrac{W - y}{\Lout} \, \fbulk, & y > W - \Lout
	\end{cases}
	\label{eq:rho}
\end{equation}
shown in Fig.~\subref{fig:setup}, leaving four parameters to optimize.

The optimal particle diameter $d^\opt$ decreases with the applied filed $B$ and $d^\opt \approx 3\xi$ for $B = 0.1\Hct$. This result is different from that in the 3D case for spherical particles, which has an optimal diameter of $d^\opt \approx 4\xi$ for the same field. This discrepancy in the result is due to the fact that the 2D circular defects we model correspond to columnar defects in 3D samples. It was found earlier that the optimal diameter of columnar defects is smaller than the optimal diameter of spherical defects by approximately one coherence length~$\xi$. Since the optimal volume fraction $f = 0.2$ and diameter of defects $d = 3\xi$ in both cases are similar,\cite{Kimmel:2017a} we keep them constant in the following analysis, making the optimization problem manageable and effectively a two parameter optimization problem.

\begin{figure}
	\vspace{-2mm}
	\includegraphics[height=166mm]{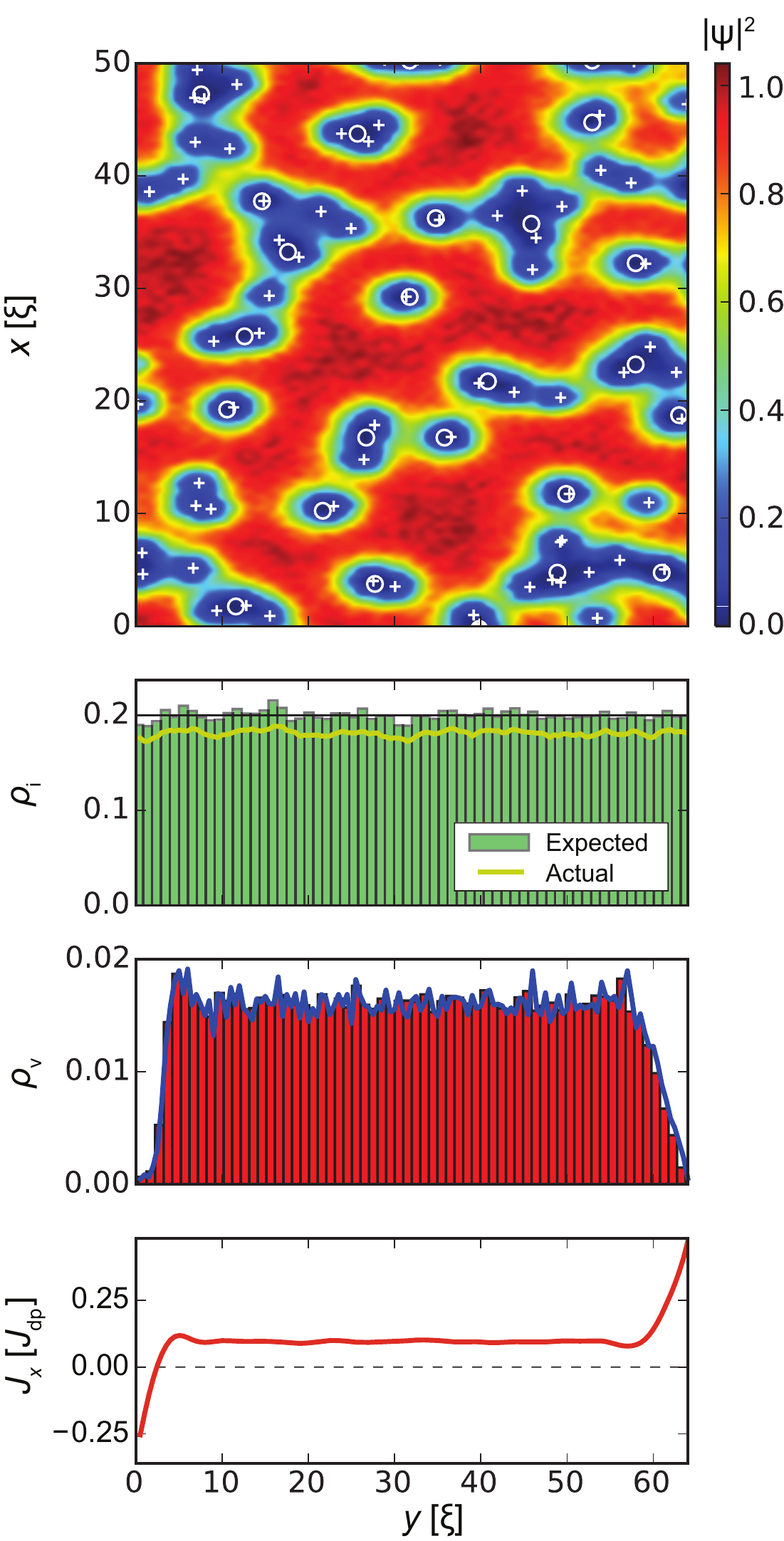}
	\vspace{-1mm}
	\caption{
		Strip with homogeneous distribution of inclusion density, 
		$\Lin = \Lout = 0$, $\rho_\mathrm{i} = f = 0.2$ in an applied 
		magnetic field $B = 0.1\Hct$.
		\textit{Top panel} shows the squared absolute value of the order 
		parameter $|\psi(\x)|^2$. Circles and crosses show inclusion 
		and vortex positions, respectively. 
		\textit{Second panel} shows the distribution of the inclusions across 
		the strip ($y$ direction). The black line shows the `requested' volume 
		fraction $f = 0.2$, the green histogram shows the distribution 
		of the centers of the inclusions, and the yellow line shows the actual 
		volume fraction occupied by the generated defects. (the actual volume 
		fraction is typically lower than the specified/requested one due to defect 
		overlaps and fluctuations of finite random number sequences.)
		\textit{Third panel} demonstrates the density $\rho_\mathrm{v}$ of vortices.
		\textit{Bottom panel} shows the local current density $J_x(y)$. 
		As expected, the edge screening currents at the surface are in opposite 
		directions, while the small local minimum and maximum a few $\xi$ 
		away from the edge are related to an alignment of vortices at the interior 
		surface barrier. The average critical current density is 
		$\Jc^\mathrm{uniform} = 0.108 \Jdp$.
	}
	\label{fig:hist_in00_out00_B01}
\end{figure}

\begin{figure}
	\vspace{-2mm}
	\includegraphics[height=166mm]{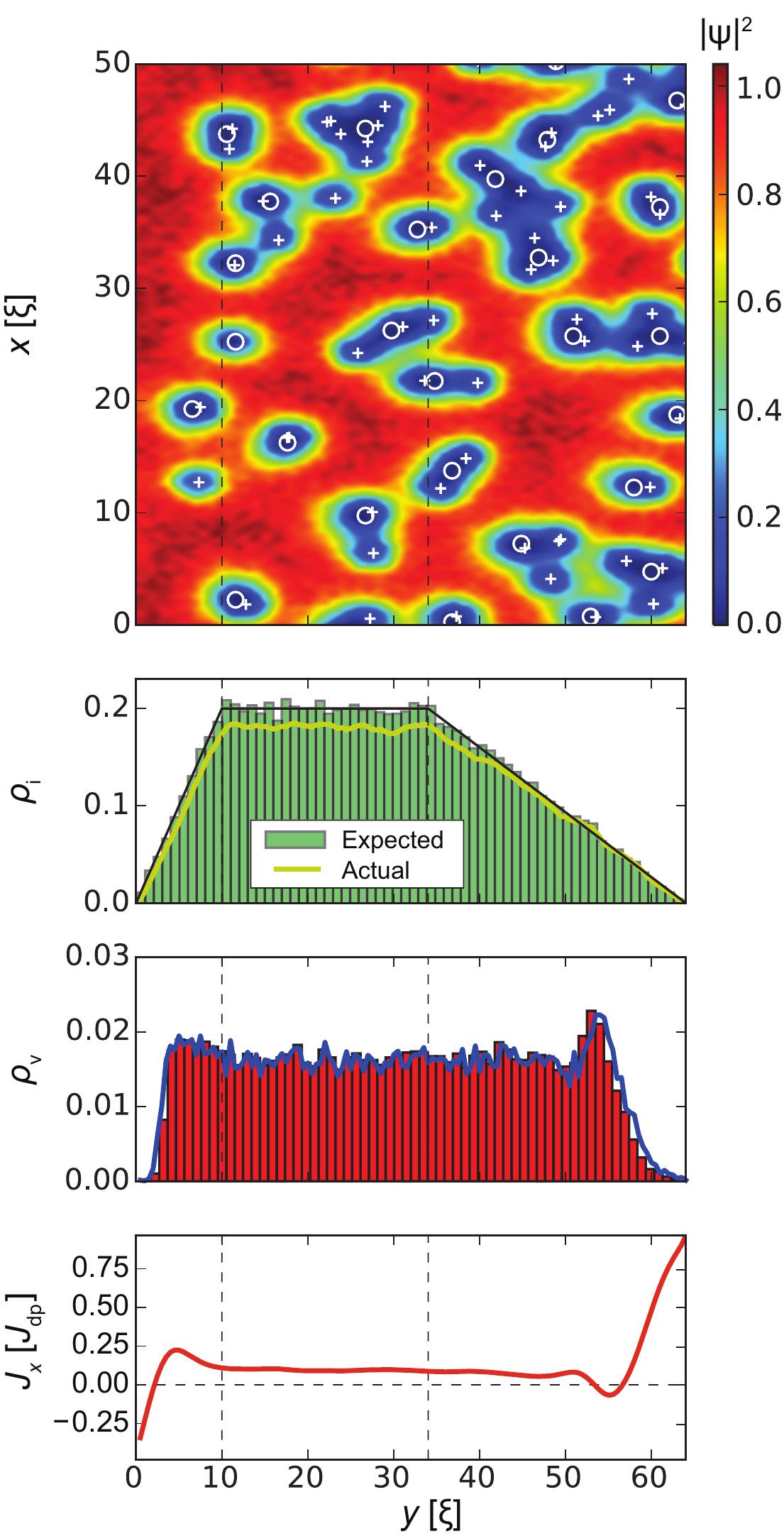}
	\vspace{-1mm}
	\caption{
		A strip with reduced inclusion density at both edges $\Lin = 10\xi$, 
		$\Lout = 30\xi$, $f = 0.2$, and $B = 0.1\Hct$. The average critical 
		current $\Jc^\mathrm{both} = 0.14 \Jdp$ is 28\% larger compared 
		to Fig.~\ref{fig:hist_in00_out00_B01}. The $J_x(y)$ dependence 
		has much more pronounced features near the edges. These 
		oscillations in the current are generated by (free) vortex rows 
		in the region of low inclusion density. 
	}
	\label{fig:hist_in10_out30_B01}
\end{figure}

\begin{figure}
	\vspace{-2mm}
	\includegraphics[height=166mm]{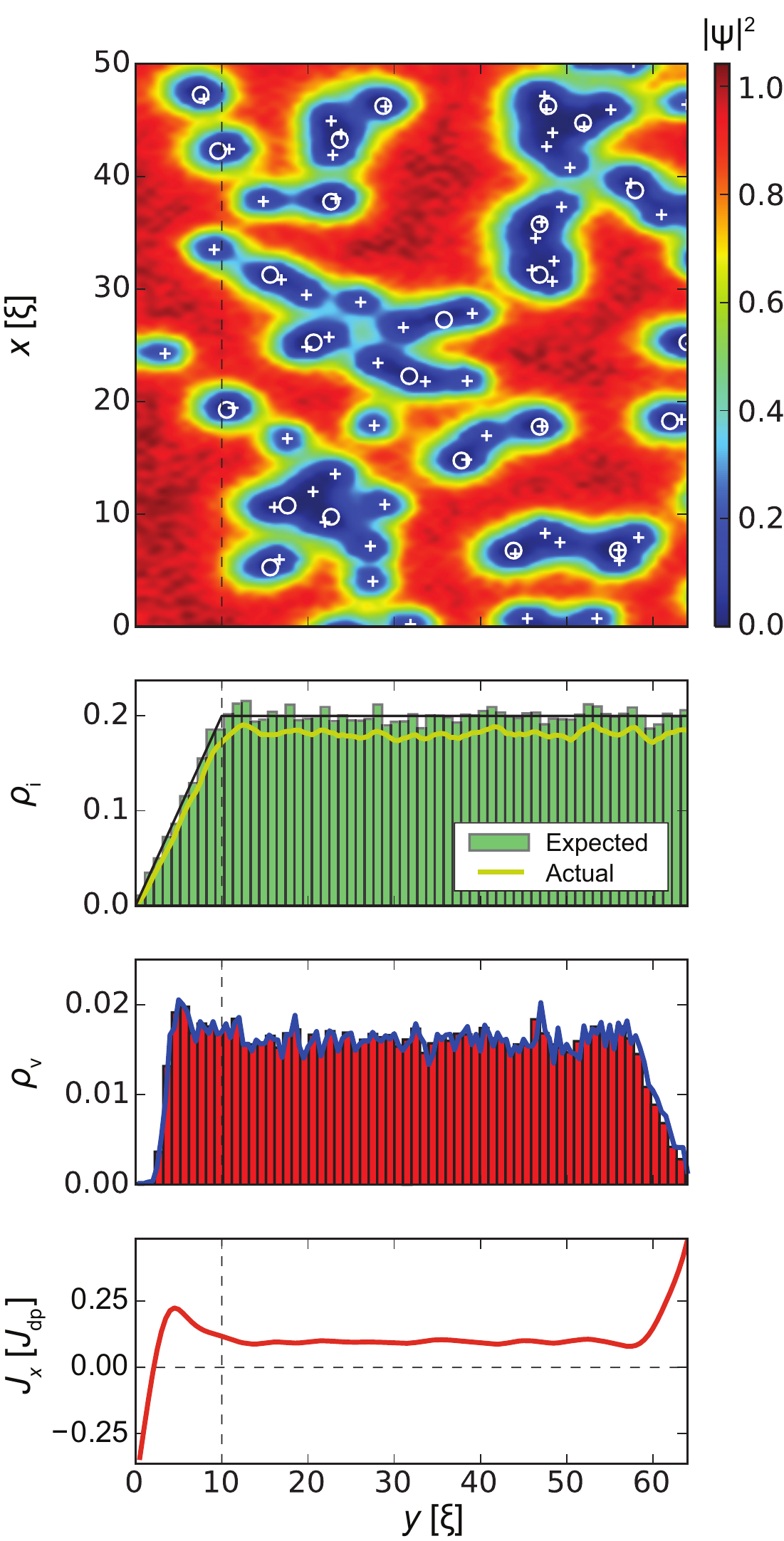}
	\vspace{-1mm}
	\caption{
		A strip with reduced inclusion density on the entrance side only, 
		$\Lin = 10 \xi$ and $\Lout = 0$, has an average critical current 
		density of $\Jc^\mathrm{in} = 0.118 \Jdp$ at applied magnetic 
		field $B = 0.1\Hct$.
	}
	\label{fig:hist_in10_out00_B01}
\end{figure}

\begin{figure}
	\vspace{-2mm}
	\includegraphics[height=166mm]{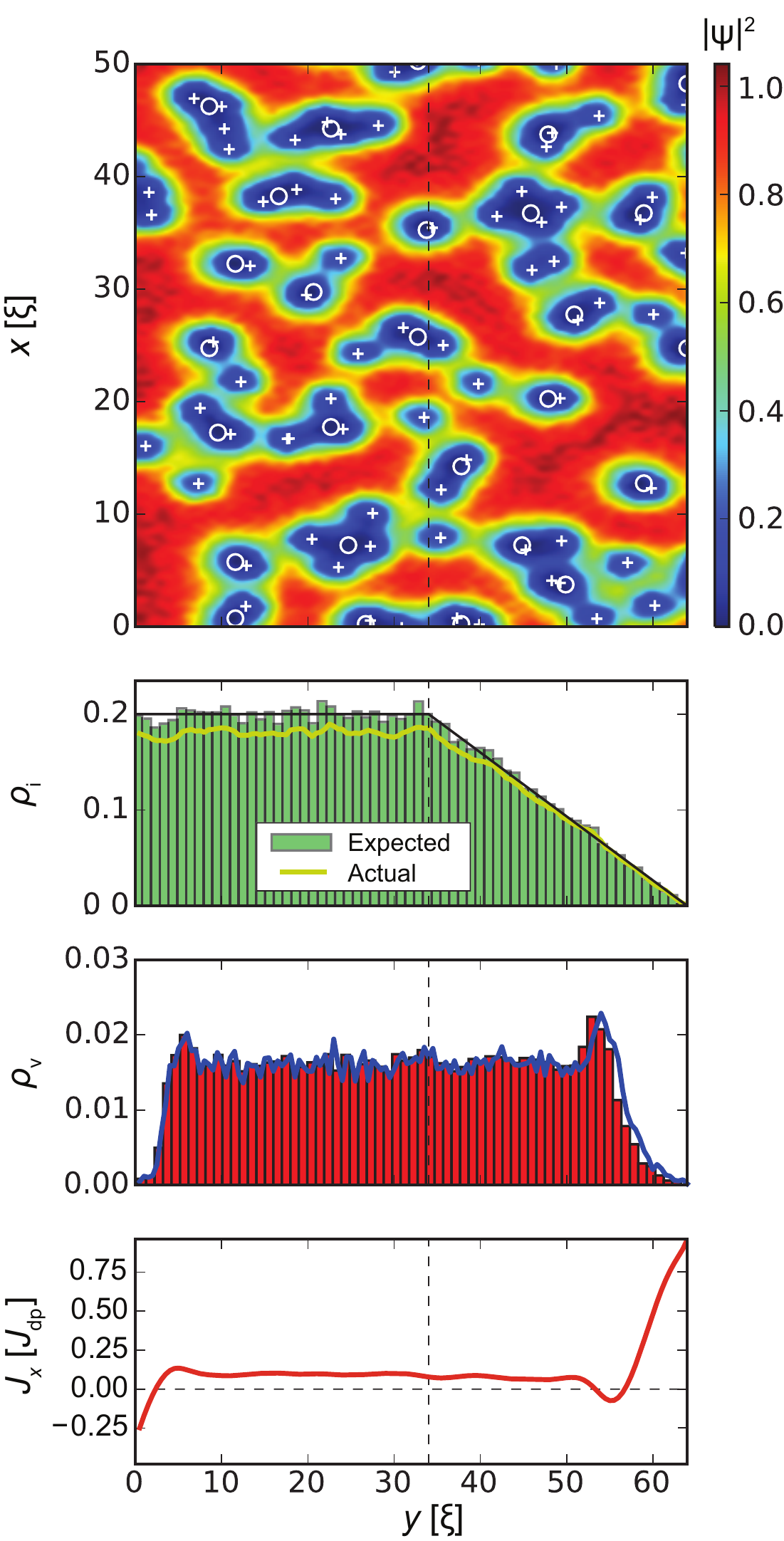}
	\vspace{-1mm}
	\caption{
		A strip with reduced inclusion density at the exit side only, 
		$\Lin = 0$ and $\Lout = 30 \xi$. The average critical current 
		density is $\Jc^\mathrm{out} = 0.131 \Jdp$ at $B = 0.1\Hct$.
	}
	\label{fig:hist_in00_out30_B01}
\end{figure}

Figure~\subref{fig:Jc_in_out} demonstrates the dependency of the critical current on the distance with reduced defect density at the entrance $\Lin$ and exit $\Lout$ of vortices for a sample of width $W = 64\xi$. One can see that the effect is far from symmetric. Figure~\subref{fig:Jc_in_out} at $B = 0.1\Hct$ shows that the critical current has a maximum of $\Jc(\Lin,\Lout) \approx 1.3 \Jc(0,0)$ at $\Lin \approx 10\xi$ and $\Lout \approx 30\xi$. The $\Jc(\Lin,\Lout)$ maxima are indicated by colored circles for $B = 0.1\Hct$, $0.2\Hct$, and $0.3\Hct$. The dependence presented in Fig.~\subref{fig:Jc_in_out} is a result of the interplay between pinning on inclusions and the Bean-Livingston barrier near the superconducting strip edge. For larger external fields the optimal entrance and exit regions become more symmetric as see by the maxima of $\Jc$ for $B = 0.2\Hct,0.2\Hct$, indicated by circles in Fig.~\subref{fig:Jc_in_out}. In particular $\Lout$  becomes smaller with increasing $B$, approaching $\Lin$, and the overall critical current peak becomes wider, i.e., the system is less sensitive to $\Lin$  and $\Lout$  at larger $B$. 

\begin{figure*}
	\vspace{-2mm}
	\subfloat{\includegraphics[height=166mm]{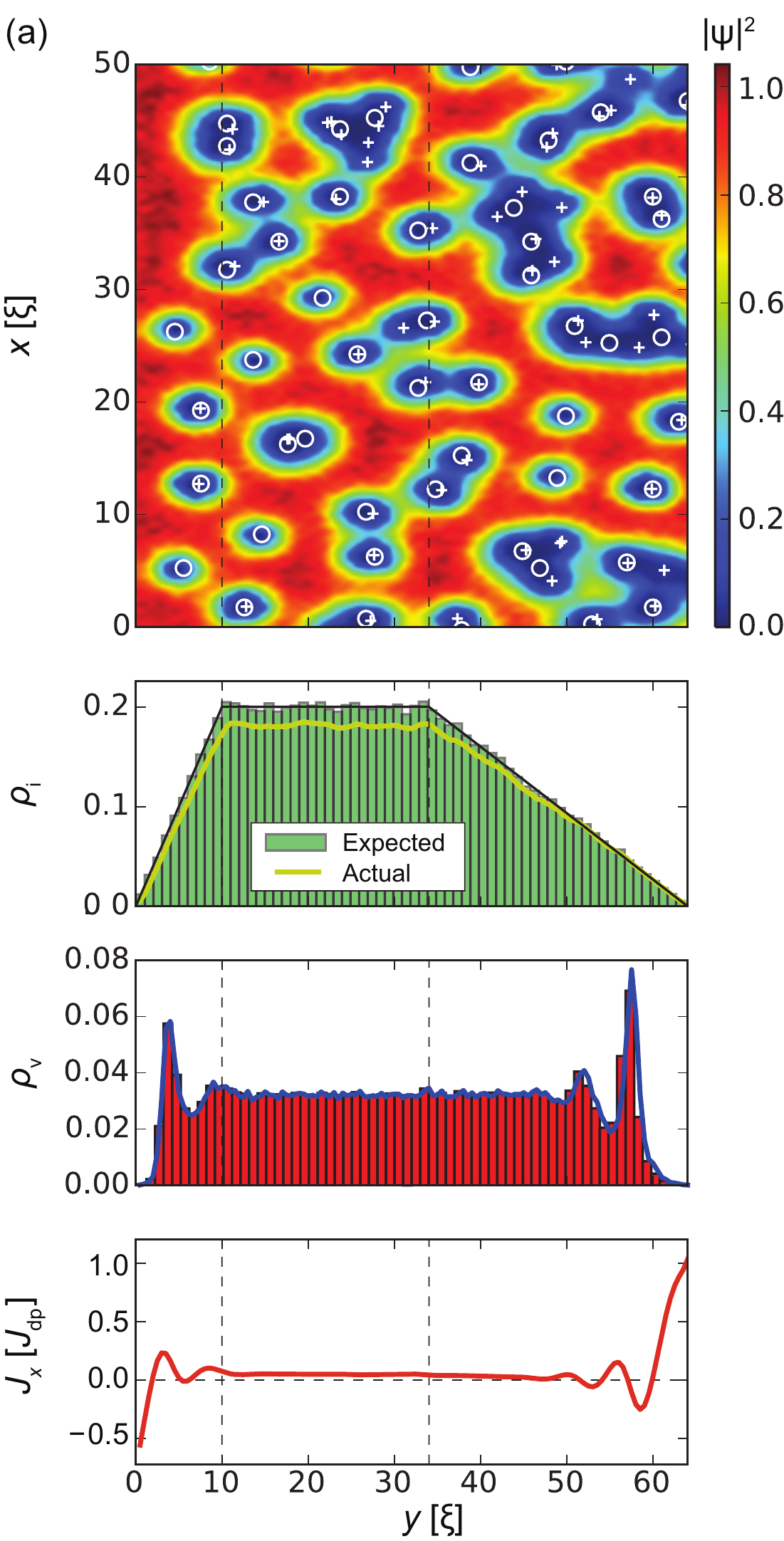}\hspace{8mm}\label{fig:hist_in10_out30_B02}}
	\subfloat{\includegraphics[height=166mm]{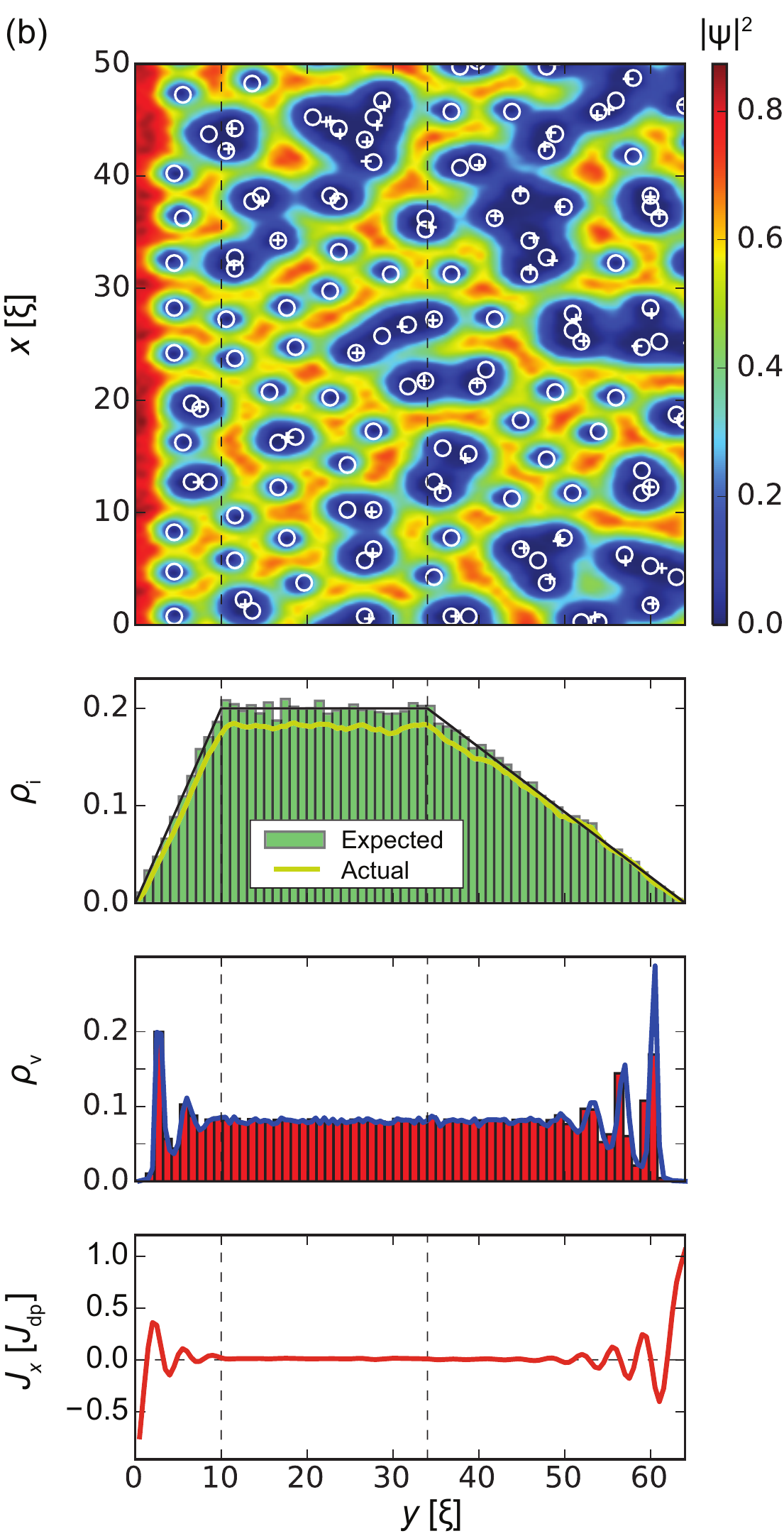}\label{fig:hist_in10_out30_B05}}
	\vspace{-1mm}
	\caption{
		The same as in Fig.~\ref{fig:hist_in10_out30_B01}, but for higher magnetic fields. 
		\subfignum{(a)}~Field $B = 0.2\Hct$ produces an average critical current density 
		$\Jc = 0.075 \Jdp$ and \subfignum{(b)}~field $B = 0.5\Hct$ generates $\Jc = 0.026 \Jdp$.
		At higher fields, vortex rows are more dense. This leads to faster oscillations in vortex 
		and local current densities $J_x(y)$. At the exit edge $J_x(y)$ reaches $\Jdp$.
	}
	\label{fig:hist_in10_out30_Bhigh}
\end{figure*}

In the following we will discuss this interplay in detail. Our results are summarized in Figs.~\ref{fig:hist_in00_out00_B01}--\ref{fig:hist_in10_out30_Bhigh}. All figures have the same format.
\textit{Top panels} show the squared absolute value of the order parameter $|\psi(\x)|^2$ in samples of width $W = 64\xi$ ($y$ direction) and length $L = 1024\xi$ ($x$ direction; quasi-periodic boundary conditions). White circles correspond to inclusions, white crosses indicate vortex positions. The presented order parameter configurations are for applied currents $J_x = \Jc$.
\textit{Second panels} show the distribution of defects along the $y$ direction and averaged over the length of the strip ($x$ direction). The black lines indicate the requested volume fraction $\rho_\mathrm{i}(y)$ defined by Eq.~\eqref{eq:rho} with $f = 0.2$ (i.e. 20\% of the volume occupied by inclusions in the bulk, which corresponds to $B_\Phi / B = 1.78$ inclusions per vortex at $B = 0.1\Hct$, where $B_\Phi = 8 f / d^2$ is the matching field of the columnar pinning landscape defined by parameters $\fbulk$ and $d$), the green histograms show the distribution of the centers of the inclusions, and the yellow lines show the actual volume fraction occupied by the generated defects. The latter value is somewhat lower than the requested value due to overlapping of inclusions and finite size effects, the real/actual volume fraction can be estimated as $\rho_\mathrm{i}^\star(y) = 1 - \exp[-\rho_\mathrm{i}(y)]$. The requested bulk defect density corresponding to a volume fraction $f = 0.2$ has $f^\star \approx 0.181$ real volume fraction. Inclusions overlapping effectively changes the matching field to $B_\Phi^\star = 8 f^\star / d^2$ and number of inclusions per one vortex to $B_\Phi^\star / B = 1.61$.
\textit{Third panels} demonstrate the density of the vortices $\rho_\mathrm{v}(y)$ averaged over the length of the strip. In all cases, the vortex density tends to zero at $y = 0$ and $y = W$ and remains roughly constant in the bulk of the superconductor.
\textit{Bottom panels} show the $x$-component of the local current density, $J_x(y)$, averaged over the length of the strip and are indicative of the edge currents and reflect the distribution of vortices.

Vortex and current density distributions for homogeneous inclusion density $\rho_\mathrm{i} = f = 0.2$ for $0 < y < W$ ($\Lin = \Lout = 0$) are shown in Fig.~\ref{fig:hist_in00_out00_B01}. The position of vortices is strongly correlated with the particular placement of the inclusions, which makes the visual analysis rather complicated. The histograms of defects, vortices, and $x$-component of current averaged over the sample length~$L$ and 10 different realizations of defect distributions contain more useful information. The vortex density is approximately constant in the bulk. This density decreases to zero at $\sim 5\xi$ away from both edges due to the Bean-Livingston barrier. Such a rapid gradient in vortex density produces large surface currents, which has a density on the order of the depairing current density $\Jdp$. The average critical current density is $\Jc^\mathrm{uniform} = 0.108 \Jdp$.

Figure~\ref{fig:hist_in10_out30_B01} shows how the result changes when we reduce the inclusion density at both edges of the superconducting strip. We pick $\Lin = 10\xi$, $\Lout = 30\xi$, with the remaining volume fraction of inclusions in the bulk as $f = 0.2$ and applied magnetic field $B = 0.1\Hct$. The chosen parameters are close to the maximum of $\Jc(\Lin,\Lout)$ shown in Fig.~\subref{fig:Jc_in_out}. The critical current $\Jc^\mathrm{both} = 0.14 \Jdp$ represents a 30\% increase compared to uniform inclusion density $\Jc^\mathrm{uniform}$. At the same time, the bulk critical current density (for $\Lin \lesssim y \lesssim L - \Lout$) remains approximately the same. This indicates that the critical current enhancement is mostly related to the defect distribution near the boundaries of the superconducting strip. 

Comparing the vortex configuration in that case with that of the uniform inclusion density case, where the location of vortices is mostly random, we find that this $\Jc$ enhancement is produced by the formation of regular vortex row(s) in the regions with a reduced concentration of defects. Each vortex row can be interpreted as an additional potential barrier parallel to the edge repelling vortices. However, since current circulates around each vortex in the row, we can observe the local current flowing in the positive $x$ direction to the right of vortex row and the current flowing in the negative $x$ direction to the left of the vortex row. The value of this local current can be as high as the depairing current density, $\Jdp$. This current density can be observed at $y = W$ in Fig.~\ref{fig:hist_in10_out30_B01}. The value of this current is somewhat lower in between rows due to cancellation of  opposite screening currents from rows at the left and at the right. Overall, these regular (mostly unpinned) rows lead to oscillations of the average vortex density and subsequently the current density along the applied current direction. This effect is similar to the one observed in artificially manufactured vortex-flow channels in irradiated mesoscopic samples.\cite{Besseling:2005} 

Next we examine how the reduced inclusion density affects the superconducting strip edge where vortices enter and exit the sample separately. The results for the strip with reduced inclusion density at the entrance side only, $\Lin = 10 \xi$ and $\Lout = 0$, is presented in Fig.~\ref{fig:hist_in10_out00_B01}. This pinning landscape generates an average critical current density $\Jc^\mathrm{in} = 0.118\Jdp$. One sees that `entrance' and bulk parts of all histograms, $y \lesssim W/2$, coincides with the corresponding part of Fig.~\ref{fig:hist_in10_out30_B01} and `exit' and bulk parts $y \gtrsim W/2$ reproduces the same regions in Fig.~\ref{fig:hist_in00_out00_B01}. An analogous situation appears with reduced inclusion density at the exit side of the strip (Fig.~\ref{fig:hist_in00_out30_B01}), $\Lin = 0$ and $\Lout = 30 \xi$. This configuration produces an average critical current density $\Jc^\mathrm{out} = 0.131 \Jdp$. 

Naturally, values of $\Jc^\mathrm{in}$ and $\Jc^\mathrm{out}$ are in between the two critical current densities of the strip with uniform inclusion distribution and the strip with reduced inclusion density on both edges, i.e., $\Jc^\mathrm{uniform} < \Jc^\mathrm{in}$, $\Jc^\mathrm{out} < \Jc^\mathrm{both}$. The independence of the vortex and current configurations on the left and right edges can also be confirmed by comparing differences in the average (or total) critical current of the four configurations discussed above. In particular, $\Jc^\mathrm{both} + \Jc^\mathrm{uniform} = \Jc^\mathrm{in} + \Jc^\mathrm{out}$ holds for all wide enough strips, $W \gtrsim \Lin + \Lout$.

Taking into account that (i) the chosen $\Lin = 10\xi$ and $\Lout = 30\xi$ correspond to the nearly largest critical current at the given magnetic field and (ii) entrance and exit edges act almost independently, we can say that the edge barrier at the entrance can generate additional critical current up to $\delta\Ic^\mathrm{in} = (\Jc^\mathrm{in} - \Jc^\mathrm{uniform})W = 0.51 \Jdp\xi$, while the same addition at the exit edge $\delta\Ic^\mathrm{out} = (\Jc^\mathrm{out} - \Jc^\mathrm{uniform})W = 1.54 \Jdp\xi$ is three times bigger. Note, that the clean strip with ideal boundaries (without any inclusions in the bulk) can generate a total critical current up to $\Ic \approx 5.1 \Jdp\xi$ at the same applied magnetic field, see Fig.~\subref{fig:Ic_w}.

Higher magnetic fields decreases the distance between neighboring vortex rows and thus leads to higher frequency oscillations of vortex density and the $x$-component of current in regions with reduced inclusion density as shown in Fig.~\ref{fig:hist_in10_out30_Bhigh}. A magnetic field $B = 0.2\Hct$ corresponds to a critical current density $\Jc = 0.075 \Jdp$ [Fig.~\subref{fig:hist_in10_out30_B02}] and field $B = 0.5\Hct$ to $\Jc = 0.026 \Jdp$ [Fig.~\subref{fig:hist_in10_out30_B05}]. On the exit side, the current density $J_x(W)$ reaches the depairing current density $\Jdp$.

\section{Discussion and conclusions} \label{sec:conclusions}

In this article we studied the interplay of surface potential barrier and bulk pinning centers in mesoscopic superconducting strips, where both pinning mechanisms are relevant. Figure~\ref{fig:Jc_Ic_w} suggests that the critical current reaches saturation at $W \sim 64\xi$ in a clean strip, meaning that the effect of the surface barriers on $\Ic$ starts to decrease above that width and bulk defects become the dominant pinning mechanism. Since non-superconducting defects are detrimental for the Bean-Livingston barrier, we studied the general case of a non-homogeneous defect distribution across the width of the strip to be able to take advantage of both mechanisms. In particular, we assumed a linear modulation of the defect concentration near both edges of the strip. This allowed us to quantify the suppression of the surface barrier by defects in the vicinity of the strip edges by studying the vortex and supercurrent distribution in these regions.

\begin{table}
	\begingroup
	\setlength{\tabcolsep}{5pt}
	\begin{tabular}{lcccccc}
		\hline \hline
		Type & $\fbulk$ & $\Lin$ & $\Lout$ & $\Jc$ & cf.\,clean & \\
		\hline
		clean & $0.0$ & -- & -- & $0.081 \Jdp$ & $100$\% & Fig.~\ref{fig:Jc_Ic_w} \\
		uniform & $0.2$ & $0$ & $0$ & $0.108 \Jdp$ & $135$\% & Fig.~\ref{fig:hist_in00_out00_B01} \\
		optimized & $0.2$ & $9\xi$ & $31\xi$ & $0.142 \Jdp$ & $178$\% & Eq.~\eqref{eq:opt} \\
		both & $0.2$ & $10\xi$ & $30\xi$ & $0.140 \Jdp$ & $175$\% & Fig.~\ref{fig:hist_in10_out30_B01} \\
		in & $0.2$ & $10\xi$ & $0$ & $0.118 \Jdp$ & $148$\% & Fig.~\ref{fig:hist_in10_out00_B01} \\
		out & $0.2$ & $0$ & $30\xi$ & $0.131 \Jdp$ & $164$\% & Fig.~\ref{fig:hist_in00_out30_B01} \\
		\hline \hline
	\end{tabular}
	\endgroup
	\caption{
		Critical currents in a strip of width $W = 64\xi$ at magnetic field $B=0.1\Hct$ 
		for different defect distributions: clean strip without defects, uniform 
		concentration of defects, optimized concentration of defects in the bulk 
		and near the edges, reduced defect concentration at both edges 
		[$\Lin, \Lout > 0$ in Eq.~\eqref{eq:rho}], and reduced defect concentration 
		at the vortex entrance and exit. The defects are circular with diameter $d = 3\xi$.
	}
\label{tab:results}
\end{table}

Table~\ref{tab:results} summarizes the results for our benchmark system --- a strip of width $W = 64\xi$ in a magnetic field $B = 0.1\Hct$. The clean strip has a critical current density of $\Jc^\mathrm{clean} = 0.081\Jdp$. For increasing strip width, the critical current density decreases as $\sim W^{-1}$, see Fig.~\ref{fig:Jc_Ic_w}, while approaching $\Jdp$ in the limit of very narrow clean strips with $W \lesssim \xi$. However, any defects or imperfections at the edges will significantly reduce these values. Adding random, but homogeneously distributed defects to the benchmark system increases $\Jc$ by $35$\% in the best case, which implies that the bulk pinning is more relevant than the suppression of the surface barrier for $B = 0.1\Hct$ and $W = 64\xi$. This maximum bulk critical current at $B = 0.1$ is reached for a volume fraction occupied by defects of $f = 0.2$ and for defects with diameter $d = 3\xi$.\cite{Kimmel:2017a} Increasing the width of these uniformly disordered strips, the effect from the edges become negligible and bulk pinning will be dominant, resulting in the critical current density approaching the one of an infinite 2D film ($\Jc^\mathrm{2D, uniform} = 0.104\Jdp$, i.e. comparable to $\Jc^\mathrm{uniform}$). Homogeneous defect distributions in narrower strips result in a noticeable suppression of the edge barrier, thus decreasing the critical current density (for $W\searrow d$ it is clear that $\Jc\to 0$).

\begin{figure}[b]
	\includegraphics[width=86mm]{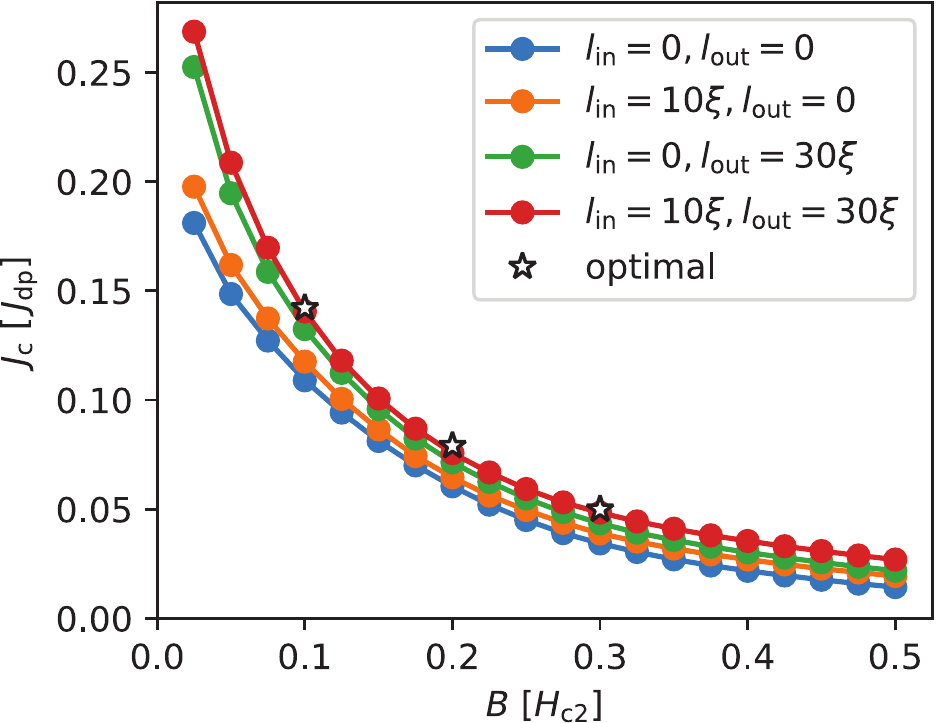}
	\caption{
		The critical current as a function of the external magnetic field for uniform 
		distribution with $\fbulk = 0.2$ (blue), `in' $\Lin = 10\xi$ (orange) and 
		`out' $\Lout = 30\xi$ (green) configurations, and `in'+`out' configuration 
		with fixed $\Lin = 10\xi$ and $\Lout = 30\xi$. The latter configuration is close 
		to the configuration (empty stars) having maximal possible $\Jc$ 
		for $B = 0.1\Hct$, $0.2\Hct$, and $0.3\Hct$.
	}
	\label{fig:Jc_B}
\end{figure}

In order to extract more detailed information about the suppression of the Bean-Livingston barrier, we introduced linear defect modulations near the edges. Studying first the vortex entrance and exit edges independently, we found that defects have an asymmetric effect on either side of the strip. A linear increase of the defect density at the entrance edge over $10\xi$ increases the critical current density by another $9$\% compared to the uniform case. A density decrease at the exit edge over $30\xi$ adds $21$\% to $\Jc$ compared to the uniform distribution. Therefore, the exit side is more sensitive to the contamination by defects located at some distance to the surface. 

Next, we studied non-uniform modulations near both edges, defined by Eq.~\eqref{eq:rho}. Subsequent optimization over its parameters $\fbulk$, $\Lin$, and $\Lout$ leads to $\Jc^\mathrm{opt} = 0.142\Jdp$, which is $31$\% more than for the uniform density with optimal values $f^\opt = 0.2$, $\Lin^\opt = 9\xi$, $\Lout^\opt = 31\xi$. Compared to the clean strip this is a $\Jc$-increase of $78$\%. We note that the effects from both sides of the strip add up independently for our relatively wide strip of $W = 64\xi$. One can expect that those optimal values for $\Lin$ and $\Lout$ remain independent of $W$ for wider strips, while their overall influence on $\Jc$ diminishes with increasing $W$ as the edges are local. Important to note is, that the mesoscopic strip under consideration with non-uniform distribution of defects has a larger critical current density than a homogeneously disordered 2D film.

Finally, we studied the field dependence of the critical current for our $W = 64\xi$ system, shown in Fig.~\ref{fig:Jc_B}. One clearly sees that the system with non-uniform defect distribution at both edge has the highest critical current density over a wide range of fields. Furthermore, as mentioned above, the system becomes less sensitive to the width of the linearly modulated edge regions as the optimal $\Jc$ value for $B = 0.2\Hct$ and $0.3\Hct$ (indicated by stars) are almost sitting on top the field dependence of the system optimized for $B = 0.1\Hct$ (red curve). Again, the homogeneously disordered system with optimal defect concentration has a lower $\Jc$ due to the suppression of the surface barrier. 

Overall, a non-homogeneous defect density modulation can significantly improve the critical current density in mesoscopic superconducting strips to higher values than those reached in 2D films.

\subsection*{Acknowledgements}

We are delighted to thank A.\,E.\,Koshelev and R.\,Willa for illuminating discussions. A.\,G. and V.\,M.\,V. were supported by the U.\,S. Department of Energy, Office of Science, Basic Energy Sciences, Materials Sciences and Engineering Division. The used code and work by G.\,K. and I.\,A.\,S were supported by the Scientific Discovery through Advanced Computing (SciDAC) program funded by U.\,S. Department of Energy, Office of Science, Advanced Scientific Computing Research and Basic Energy Science, Division of Materials Science and Engineering. 

Simulations were performed at \href{https://www.olcf.ornl.gov/}{Oak Ridge LCF} supported by DOE under contract DE-AC05-00OR22725, at \href{https://www.alcf.anl.gov/}{Argonne LCF} (DOE contract DE-AC02-06CH11357), and \href{http://www.cs.niu.edu/crcd/}{Computing Facility} at Northern Illinois University. Other simulations using time-dependent Ginzburg-Landau model can be found at \href{http://oscon-scidac.org/}{OSCon website} and \href{https://www.youtube.com/channel/UCjdQ4Ruhxma5pkxGrFxw3CA}{YouTube channel}.

\bibliography{edge_effect}

\begin{thebibliography}{40}%
\makeatletter
\providecommand \@ifxundefined [1]{%
 \@ifx{#1\undefined}
}%
\providecommand \@ifnum [1]{%
 \ifnum #1\expandafter \@firstoftwo
 \else \expandafter \@secondoftwo
 \fi
}%
\providecommand \@ifx [1]{%
 \ifx #1\expandafter \@firstoftwo
 \else \expandafter \@secondoftwo
 \fi
}%
\providecommand \natexlab [1]{#1}%
\providecommand \enquote  [1]{``#1''}%
\providecommand \bibnamefont  [1]{#1}%
\providecommand \bibfnamefont [1]{#1}%
\providecommand \citenamefont [1]{#1}%
\providecommand \href@noop [0]{\@secondoftwo}%
\providecommand \href [0]{\begingroup \@sanitize@url \@href}%
\providecommand \@href[1]{\@@startlink{#1}\@@href}%
\providecommand \@@href[1]{\endgroup#1\@@endlink}%
\providecommand \@sanitize@url [0]{\catcode `\\12\catcode `\$12\catcode
  `\&12\catcode `\#12\catcode `\^12\catcode `\_12\catcode `\%12\relax}%
\providecommand \@@startlink[1]{}%
\providecommand \@@endlink[0]{}%
\providecommand \url  [0]{\begingroup\@sanitize@url \@url }%
\providecommand \@url [1]{\endgroup\@href {#1}{\urlprefix }}%
\providecommand \urlprefix  [0]{URL }%
\providecommand \Eprint [0]{\href }%
\providecommand \doibase [0]{http://dx.doi.org/}%
\providecommand \selectlanguage [0]{\@gobble}%
\providecommand \bibinfo  [0]{\@secondoftwo}%
\providecommand \bibfield  [0]{\@secondoftwo}%
\providecommand \translation [1]{[#1]}%
\providecommand \BibitemOpen [0]{}%
\providecommand \bibitemStop [0]{}%
\providecommand \bibitemNoStop [0]{.\EOS\space}%
\providecommand \EOS [0]{\spacefactor3000\relax}%
\providecommand \BibitemShut  [1]{\csname bibitem#1\endcsname}%
\let\auto@bib@innerbib\@empty
\bibitem [{\citenamefont {Blatter}\ \emph {et~al.}(1994)\citenamefont
  {Blatter}, \citenamefont {Feigel'man}, \citenamefont {Geshkenbein},
  \citenamefont {Larkin},\ and\ \citenamefont {Vinokur}}]{Blatter:1994}%
  \BibitemOpen
  \bibfield  {author} {\bibinfo {author} {\bibfnamefont {G.}~\bibnamefont
  {Blatter}}, \bibinfo {author} {\bibfnamefont {M.~V.}\ \bibnamefont
  {Feigel'man}}, \bibinfo {author} {\bibfnamefont {V.~B.}\ \bibnamefont
  {Geshkenbein}}, \bibinfo {author} {\bibfnamefont {A.~I.}\ \bibnamefont
  {Larkin}}, \ and\ \bibinfo {author} {\bibfnamefont {V.~M.}\ \bibnamefont
  {Vinokur}},\ }\bibfield  {title} {\emph {\bibinfo {title} {Vortices in
  high-temperature superconductors}},\ }\href {\doibase
  10.1103/RevModPhys.66.1125} {\bibfield  {journal} {\bibinfo  {journal} {Rev.
  Mod. Phys.}\ }\textbf {\bibinfo {volume} {66}},\ \bibinfo {pages} {1125}
  (\bibinfo {year} {1994})}\BibitemShut {NoStop}%
\bibitem [{\citenamefont {Holesinger}\ \emph {et~al.}(2008)\citenamefont
  {Holesinger}, \citenamefont {Civale}, \citenamefont {Maiorov}, \citenamefont
  {Feldmann}, \citenamefont {Coulter}, \citenamefont {Miller}, \citenamefont
  {Maroni}, \citenamefont {Chen}, \citenamefont {Larbalestier}, \citenamefont
  {Feenstra}, \citenamefont {Li}, \citenamefont {Huang}, \citenamefont
  {Kodenkandath}, \citenamefont {Zhang}, \citenamefont {Rupich},\ and\
  \citenamefont {Malozemoff}}]{Holesinger:2008}%
  \BibitemOpen
  \bibfield  {author} {\bibinfo {author} {\bibfnamefont {T.~G.}\ \bibnamefont
  {Holesinger}}, \bibinfo {author} {\bibfnamefont {L.}~\bibnamefont {Civale}},
  \bibinfo {author} {\bibfnamefont {B.}~\bibnamefont {Maiorov}}, \bibinfo
  {author} {\bibfnamefont {D.~M.}\ \bibnamefont {Feldmann}}, \bibinfo {author}
  {\bibfnamefont {J.~Y.}\ \bibnamefont {Coulter}}, \bibinfo {author}
  {\bibfnamefont {D.~J.}\ \bibnamefont {Miller}}, \bibinfo {author}
  {\bibfnamefont {V.~A.}\ \bibnamefont {Maroni}}, \bibinfo {author}
  {\bibfnamefont {Z.}~\bibnamefont {Chen}}, \bibinfo {author} {\bibfnamefont
  {D.~C.}\ \bibnamefont {Larbalestier}}, \bibinfo {author} {\bibfnamefont
  {R.}~\bibnamefont {Feenstra}}, \bibinfo {author} {\bibfnamefont
  {X.}~\bibnamefont {Li}}, \bibinfo {author} {\bibfnamefont {Y.}~\bibnamefont
  {Huang}}, \bibinfo {author} {\bibfnamefont {T.}~\bibnamefont {Kodenkandath}},
  \bibinfo {author} {\bibfnamefont {W.}~\bibnamefont {Zhang}}, \bibinfo
  {author} {\bibfnamefont {M.}~\bibnamefont {Rupich}}, \ and\ \bibinfo {author}
  {\bibfnamefont {A.}~\bibnamefont {Malozemoff}},\ }\bibfield  {title} {\emph
  {\bibinfo {title} {Progress in nanoengineered microstructures for tunable
  high-current, high-temperature superconducting wires}},\ }\href {\doibase
  10.1002/adma.200700919} {\bibfield  {journal} {\bibinfo  {journal} {Adv.
  Mater.}\ }\textbf {\bibinfo {volume} {20}},\ \bibinfo {pages} {391} (\bibinfo
  {year} {2008})}\BibitemShut {NoStop}%
\bibitem [{\citenamefont {Malozemoff}(2012)}]{Malozemoff:2012}%
  \BibitemOpen
  \bibfield  {author} {\bibinfo {author} {\bibfnamefont {A.~P.}\ \bibnamefont
  {Malozemoff}},\ }\bibfield  {title} {\emph {\bibinfo {title}
  {Second-generation high-temperature superconductor wires for the electric
  power grid}},\ }\href {\doibase 10.1146/annurev-matsci-100511-100240}
  {\bibfield  {journal} {\bibinfo  {journal} {Annu. Rev. Mater. Res.}\ }\textbf
  {\bibinfo {volume} {42}},\ \bibinfo {pages} {373} (\bibinfo {year}
  {2012})}\BibitemShut {NoStop}%
\bibitem [{\citenamefont {Kwok}\ \emph {et~al.}(2016)\citenamefont {Kwok},
  \citenamefont {Welp}, \citenamefont {Glatz}, \citenamefont {Koshelev},
  \citenamefont {Kihlstrom},\ and\ \citenamefont {Crabtree}}]{Kwok:2016}%
  \BibitemOpen
  \bibfield  {author} {\bibinfo {author} {\bibfnamefont {W.-K.}\ \bibnamefont
  {Kwok}}, \bibinfo {author} {\bibfnamefont {U.}~\bibnamefont {Welp}}, \bibinfo
  {author} {\bibfnamefont {A.}~\bibnamefont {Glatz}}, \bibinfo {author}
  {\bibfnamefont {A.~E.}\ \bibnamefont {Koshelev}}, \bibinfo {author}
  {\bibfnamefont {K.~J.}\ \bibnamefont {Kihlstrom}}, \ and\ \bibinfo {author}
  {\bibfnamefont {G.~W.}\ \bibnamefont {Crabtree}},\ }\bibfield  {title} {\emph
  {\bibinfo {title} {Vortices in high-performance high-temperature
  superconductors}},\ }\href {\doibase 10.1088/0034-4885/79/11/116501}
  {\bibfield  {journal} {\bibinfo  {journal} {Rep. Prog. Phys.}\ }\textbf
  {\bibinfo {volume} {79}},\ \bibinfo {pages} {116501} (\bibinfo {year}
  {2016})}\BibitemShut {NoStop}%
\bibitem [{\citenamefont {Stan}\ \emph {et~al.}(2004)\citenamefont {Stan},
  \citenamefont {Field},\ and\ \citenamefont {Martinis}}]{Stan:2004}%
  \BibitemOpen
  \bibfield  {author} {\bibinfo {author} {\bibfnamefont {G.}~\bibnamefont
  {Stan}}, \bibinfo {author} {\bibfnamefont {S.~B.}\ \bibnamefont {Field}}, \
  and\ \bibinfo {author} {\bibfnamefont {J.~M.}\ \bibnamefont {Martinis}},\
  }\bibfield  {title} {\emph {\bibinfo {title} {Critical field for complete
  vortex expulsion from narrow superconducting strips}},\ }\href {\doibase
  10.1103/PhysRevLett.92.097003} {\bibfield  {journal} {\bibinfo  {journal}
  {Phys. Rev. Lett.}\ }\textbf {\bibinfo {volume} {92}},\ \bibinfo {pages}
  {097003} (\bibinfo {year} {2004})}\BibitemShut {NoStop}%
\bibitem [{\citenamefont {Zeldov}\ \emph {et~al.}(1994)\citenamefont {Zeldov},
  \citenamefont {Larkin}, \citenamefont {Geshkenbein}, \citenamefont
  {Konczykowski}, \citenamefont {Majer}, \citenamefont {Khaykovich},
  \citenamefont {Vinokur},\ and\ \citenamefont {Shtrikman}}]{Zeldov:1994}%
  \BibitemOpen
  \bibfield  {author} {\bibinfo {author} {\bibfnamefont {E.}~\bibnamefont
  {Zeldov}}, \bibinfo {author} {\bibfnamefont {A.~I.}\ \bibnamefont {Larkin}},
  \bibinfo {author} {\bibfnamefont {V.~B.}\ \bibnamefont {Geshkenbein}},
  \bibinfo {author} {\bibfnamefont {M.}~\bibnamefont {Konczykowski}}, \bibinfo
  {author} {\bibfnamefont {D.}~\bibnamefont {Majer}}, \bibinfo {author}
  {\bibfnamefont {B.}~\bibnamefont {Khaykovich}}, \bibinfo {author}
  {\bibfnamefont {V.~M.}\ \bibnamefont {Vinokur}}, \ and\ \bibinfo {author}
  {\bibfnamefont {H.}~\bibnamefont {Shtrikman}},\ }\bibfield  {title} {\emph
  {\bibinfo {title} {Geometrical barriers in high-temperature
  superconductors}},\ }\href {\doibase 10.1103/PhysRevLett.73.1428} {\bibfield
  {journal} {\bibinfo  {journal} {Phys. Rev. Lett.}\ }\textbf {\bibinfo
  {volume} {73}},\ \bibinfo {pages} {1428} (\bibinfo {year}
  {1994})}\BibitemShut {NoStop}%
\bibitem [{\citenamefont {Kuit}\ \emph {et~al.}(2008)\citenamefont {Kuit},
  \citenamefont {Kirtley}, \citenamefont {van~der Veur}, \citenamefont
  {Molenaar}, \citenamefont {Roesthuis}, \citenamefont {Troeman}, \citenamefont
  {Clem}, \citenamefont {Hilgenkamp}, \citenamefont {Rogalla},\ and\
  \citenamefont {Flokstra}}]{Kuit:2008}%
  \BibitemOpen
  \bibfield  {author} {\bibinfo {author} {\bibfnamefont {K.~H.}\ \bibnamefont
  {Kuit}}, \bibinfo {author} {\bibfnamefont {J.~R.}\ \bibnamefont {Kirtley}},
  \bibinfo {author} {\bibfnamefont {W.}~\bibnamefont {van~der Veur}}, \bibinfo
  {author} {\bibfnamefont {C.~G.}\ \bibnamefont {Molenaar}}, \bibinfo {author}
  {\bibfnamefont {F.~J.~G.}\ \bibnamefont {Roesthuis}}, \bibinfo {author}
  {\bibfnamefont {A.~G.~P.}\ \bibnamefont {Troeman}}, \bibinfo {author}
  {\bibfnamefont {J.~R.}\ \bibnamefont {Clem}}, \bibinfo {author}
  {\bibfnamefont {H.}~\bibnamefont {Hilgenkamp}}, \bibinfo {author}
  {\bibfnamefont {H.}~\bibnamefont {Rogalla}}, \ and\ \bibinfo {author}
  {\bibfnamefont {J.}~\bibnamefont {Flokstra}},\ }\bibfield  {title} {\emph
  {\bibinfo {title} {Vortex trapping and expulsion in thin-film
  {YBa$_2$Cu$_3$O$_{7-\delta}$} strips}},\ }\href {\doibase
  10.1103/PhysRevB.77.134504} {\bibfield  {journal} {\bibinfo  {journal} {Phys.
  Rev. B}\ }\textbf {\bibinfo {volume} {77}},\ \bibinfo {pages} {134504}
  (\bibinfo {year} {2008})}\BibitemShut {NoStop}%
\bibitem [{\citenamefont {Vodolazov}(2013)}]{Vodolazov:2013}%
  \BibitemOpen
  \bibfield  {author} {\bibinfo {author} {\bibfnamefont {D.~Y.}\ \bibnamefont
  {Vodolazov}},\ }\bibfield  {title} {\emph {\bibinfo {title} {Vortex-induced
  negative magnetoresistance and peak effect in narrow superconducting
  films}},\ }\href {\doibase 10.1103/PhysRevB.88.014525} {\bibfield  {journal}
  {\bibinfo  {journal} {Phys. Rev. B}\ }\textbf {\bibinfo {volume} {88}},\
  \bibinfo {pages} {014525} (\bibinfo {year} {2013})}\BibitemShut {NoStop}%
\bibitem [{\citenamefont {Willa}\ \emph {et~al.}(2014)\citenamefont {Willa},
  \citenamefont {Geshkenbein},\ and\ \citenamefont {Blatter}}]{Willa:2014}%
  \BibitemOpen
  \bibfield  {author} {\bibinfo {author} {\bibfnamefont {R.}~\bibnamefont
  {Willa}}, \bibinfo {author} {\bibfnamefont {V.~B.}\ \bibnamefont
  {Geshkenbein}}, \ and\ \bibinfo {author} {\bibfnamefont {G.}~\bibnamefont
  {Blatter}},\ }\bibfield  {title} {\emph {\bibinfo {title} {Suppression of
  geometric barrier in {type-II} superconducting strips}},\ }\href {\doibase
  10.1103/PhysRevB.89.104514} {\bibfield  {journal} {\bibinfo  {journal} {Phys.
  Rev. B}\ }\textbf {\bibinfo {volume} {89}},\ \bibinfo {pages} {104514}
  (\bibinfo {year} {2014})}\BibitemShut {NoStop}%
\bibitem [{\citenamefont {Papari}\ \emph {et~al.}(2016)\citenamefont {Papari},
  \citenamefont {Glatz}, \citenamefont {Carillo}, \citenamefont {Stornaiuolo},
  \citenamefont {Massarotti}, \citenamefont {Rouco}, \citenamefont
  {Longobardi}, \citenamefont {Beltram}, \citenamefont {Vinokur},\ and\
  \citenamefont {Tafuri}}]{Papari:2016}%
  \BibitemOpen
  \bibfield  {author} {\bibinfo {author} {\bibfnamefont {G.~P.}\ \bibnamefont
  {Papari}}, \bibinfo {author} {\bibfnamefont {A.}~\bibnamefont {Glatz}},
  \bibinfo {author} {\bibfnamefont {F.}~\bibnamefont {Carillo}}, \bibinfo
  {author} {\bibfnamefont {D.}~\bibnamefont {Stornaiuolo}}, \bibinfo {author}
  {\bibfnamefont {D.}~\bibnamefont {Massarotti}}, \bibinfo {author}
  {\bibfnamefont {V.}~\bibnamefont {Rouco}}, \bibinfo {author} {\bibfnamefont
  {L.}~\bibnamefont {Longobardi}}, \bibinfo {author} {\bibfnamefont
  {F.}~\bibnamefont {Beltram}}, \bibinfo {author} {\bibfnamefont {V.~M.}\
  \bibnamefont {Vinokur}}, \ and\ \bibinfo {author} {\bibfnamefont
  {F.}~\bibnamefont {Tafuri}},\ }\bibfield  {title} {\emph {\bibinfo {title}
  {Geometrical vortex lattice pinning and melting in {YBaCuO} submicron
  bridges}},\ }\href {\doibase doi:10.1038/srep38677} {\bibfield  {journal}
  {\bibinfo  {journal} {Sci. Rep.}\ }\textbf {\bibinfo {volume} {6}},\ \bibinfo
  {pages} {38677} (\bibinfo {year} {2016})}\BibitemShut {NoStop}%
\bibitem [{\citenamefont {Wang}\ \emph {et~al.}(2017)\citenamefont {Wang},
  \citenamefont {Glatz}, \citenamefont {Kimmel}, \citenamefont {Aranson},
  \citenamefont {Thoutam}, \citenamefont {Xiao}, \citenamefont {Berdiyorov},
  \citenamefont {Peeters}, \citenamefont {Crabtree},\ and\ \citenamefont
  {Kwok}}]{Wang:2017}%
  \BibitemOpen
  \bibfield  {author} {\bibinfo {author} {\bibfnamefont {Y.-L.}\ \bibnamefont
  {Wang}}, \bibinfo {author} {\bibfnamefont {A.}~\bibnamefont {Glatz}},
  \bibinfo {author} {\bibfnamefont {G.~J.}\ \bibnamefont {Kimmel}}, \bibinfo
  {author} {\bibfnamefont {I.~S.}\ \bibnamefont {Aranson}}, \bibinfo {author}
  {\bibfnamefont {L.~R.}\ \bibnamefont {Thoutam}}, \bibinfo {author}
  {\bibfnamefont {Z.-L.}\ \bibnamefont {Xiao}}, \bibinfo {author}
  {\bibfnamefont {G.~R.}\ \bibnamefont {Berdiyorov}}, \bibinfo {author}
  {\bibfnamefont {F.~M.}\ \bibnamefont {Peeters}}, \bibinfo {author}
  {\bibfnamefont {G.~W.}\ \bibnamefont {Crabtree}}, \ and\ \bibinfo {author}
  {\bibfnamefont {W.-K.}\ \bibnamefont {Kwok}},\ }\bibfield  {title} {\emph
  {\bibinfo {title} {Parallel magnetic field suppresses dissipation in
  superconducting nanostrips}},\ }\href {\doibase 10.1073/pnas.1619550114}
  {\bibfield  {journal} {\bibinfo  {journal} {Proc. Natl. Acad. Sci.}\ }\textbf
  {\bibinfo {volume} {114}},\ \bibinfo {pages} {E10274} (\bibinfo {year}
  {2017})}\BibitemShut {NoStop}%
\bibitem [{\citenamefont {Sadovskyy}\ \emph {et~al.}(2018)\citenamefont
  {Sadovskyy}, \citenamefont {Koshelev}, \citenamefont {Kwok}, \citenamefont
  {Welp},\ and\ \citenamefont {Glatz}}]{Sadovskyy:2018}%
  \BibitemOpen
  \bibfield  {author} {\bibinfo {author} {\bibfnamefont {I.~A.}\ \bibnamefont
  {Sadovskyy}}, \bibinfo {author} {\bibfnamefont {A.~E.}\ \bibnamefont
  {Koshelev}}, \bibinfo {author} {\bibfnamefont {W.-K.}\ \bibnamefont {Kwok}},
  \bibinfo {author} {\bibfnamefont {U.}~\bibnamefont {Welp}}, \ and\ \bibinfo
  {author} {\bibfnamefont {A.}~\bibnamefont {Glatz}},\ }\bibfield  {title}
  {\emph {\bibinfo {title} {Targeted evolution of pinning landscapes for large
  critical currents}},\ }\href@noop {} {\bibfield  {journal} {\bibinfo
  {journal} {submitted}\ } (\bibinfo {year} {2018})}\BibitemShut {NoStop}%
\bibitem [{\citenamefont {C{\'o}rdoba}\ \emph {et~al.}(2013)\citenamefont
  {C{\'o}rdoba}, \citenamefont {Baturina}, \citenamefont {Ses{\'e}},
  \citenamefont {Yu~Mironov}, \citenamefont {De~Teresa}, \citenamefont
  {Ibarra}, \citenamefont {Nasimov}, \citenamefont {Gutakovskii}, \citenamefont
  {Latyshev}, \citenamefont {Guillam{\'o}n}, \citenamefont {Suderow},
  \citenamefont {Vieira}, \citenamefont {Baklanov}, \citenamefont {Palacios},\
  and\ \citenamefont {Vinokur}}]{Cordoba:2013}%
  \BibitemOpen
  \bibfield  {author} {\bibinfo {author} {\bibfnamefont {R.}~\bibnamefont
  {C{\'o}rdoba}}, \bibinfo {author} {\bibfnamefont {T.~I.}\ \bibnamefont
  {Baturina}}, \bibinfo {author} {\bibfnamefont {J.}~\bibnamefont {Ses{\'e}}},
  \bibinfo {author} {\bibfnamefont {A.}~\bibnamefont {Yu~Mironov}}, \bibinfo
  {author} {\bibfnamefont {J.~M.}\ \bibnamefont {De~Teresa}}, \bibinfo {author}
  {\bibfnamefont {M.~R.}\ \bibnamefont {Ibarra}}, \bibinfo {author}
  {\bibfnamefont {D.~A.}\ \bibnamefont {Nasimov}}, \bibinfo {author}
  {\bibfnamefont {A.~K.}\ \bibnamefont {Gutakovskii}}, \bibinfo {author}
  {\bibfnamefont {A.~V.}\ \bibnamefont {Latyshev}}, \bibinfo {author}
  {\bibfnamefont {I.}~\bibnamefont {Guillam{\'o}n}}, \bibinfo {author}
  {\bibfnamefont {H.}~\bibnamefont {Suderow}}, \bibinfo {author} {\bibfnamefont
  {S.}~\bibnamefont {Vieira}}, \bibinfo {author} {\bibfnamefont {M.~R.}\
  \bibnamefont {Baklanov}}, \bibinfo {author} {\bibfnamefont {J.~J.}\
  \bibnamefont {Palacios}}, \ and\ \bibinfo {author} {\bibfnamefont {V.~M.}\
  \bibnamefont {Vinokur}},\ }\bibfield  {title} {\emph {\bibinfo {title}
  {Magnetic field-induced dissipation-free state in superconducting
  nanostructures}},\ }\href {\doibase 10.1038/ncomms2437} {\bibfield  {journal}
  {\bibinfo  {journal} {Nat. Commun.}\ }\textbf {\bibinfo {volume} {4}},\
  \bibinfo {pages} {1437} (\bibinfo {year} {2013})}\BibitemShut {NoStop}%
\bibitem [{\citenamefont {Berdiyorov}\ \emph {et~al.}(2012)\citenamefont
  {Berdiyorov}, \citenamefont {Milo\v{s}evi\'{c}}, \citenamefont {Latimer},
  \citenamefont {Xiao}, \citenamefont {Kwok},\ and\ \citenamefont
  {Peeters}}]{Berdiyorov:2012}%
  \BibitemOpen
  \bibfield  {author} {\bibinfo {author} {\bibfnamefont {G.~R.}\ \bibnamefont
  {Berdiyorov}}, \bibinfo {author} {\bibfnamefont {M.~V.}\ \bibnamefont
  {Milo\v{s}evi\'{c}}}, \bibinfo {author} {\bibfnamefont {M.~L.}\ \bibnamefont
  {Latimer}}, \bibinfo {author} {\bibfnamefont {Z.~L.}\ \bibnamefont {Xiao}},
  \bibinfo {author} {\bibfnamefont {W.~K.}\ \bibnamefont {Kwok}}, \ and\
  \bibinfo {author} {\bibfnamefont {F.~M.}\ \bibnamefont {Peeters}},\
  }\bibfield  {title} {\emph {\bibinfo {title} {Large magnetoresistance
  oscillations in mesoscopic superconductors due to current-excited moving
  vortices}},\ }\href {\doibase 10.1103/PhysRevLett.109.057004} {\bibfield
  {journal} {\bibinfo  {journal} {Phys. Rev. Lett.}\ }\textbf {\bibinfo
  {volume} {109}},\ \bibinfo {pages} {057004} (\bibinfo {year}
  {2012})}\BibitemShut {NoStop}%
\bibitem [{\citenamefont {Kupriyanov}\ and\ \citenamefont
  {Likharev}(1975)}]{Kupriyanov:1974}%
  \BibitemOpen
  \bibfield  {author} {\bibinfo {author} {\bibfnamefont {M.~Y.}\ \bibnamefont
  {Kupriyanov}}\ and\ \bibinfo {author} {\bibfnamefont {K.~K.}\ \bibnamefont
  {Likharev}},\ }\bibfield  {title} {\emph {\bibinfo {title} {Effect of an edge
  barrier on the critical current of a superconducting film}},\ }\href@noop {}
  {\bibfield  {journal} {\bibinfo  {journal} {Sov. Phys. Solid State}\ }\textbf
  {\bibinfo {volume} {16}},\ \bibinfo {pages} {1835} (\bibinfo {year}
  {1975})}\BibitemShut {NoStop}%
\bibitem [{\citenamefont {Tahara}\ \emph {et~al.}(1990)\citenamefont {Tahara},
  \citenamefont {Anlage}, \citenamefont {Halbritter}, \citenamefont {Eom},
  \citenamefont {Fork}, \citenamefont {Geballe},\ and\ \citenamefont
  {Beasley}}]{Tahara:1990}%
  \BibitemOpen
  \bibfield  {author} {\bibinfo {author} {\bibfnamefont {S.}~\bibnamefont
  {Tahara}}, \bibinfo {author} {\bibfnamefont {S.~M.}\ \bibnamefont {Anlage}},
  \bibinfo {author} {\bibfnamefont {J.}~\bibnamefont {Halbritter}}, \bibinfo
  {author} {\bibfnamefont {C.-B.}\ \bibnamefont {Eom}}, \bibinfo {author}
  {\bibfnamefont {D.~K.}\ \bibnamefont {Fork}}, \bibinfo {author}
  {\bibfnamefont {T.~H.}\ \bibnamefont {Geballe}}, \ and\ \bibinfo {author}
  {\bibfnamefont {M.~R.}\ \bibnamefont {Beasley}},\ }\bibfield  {title} {\emph
  {\bibinfo {title} {Critical currents, pinning, and edge barriers in narrow
  {YBa$_2$Cu$_3$O$_{7-\delta}$} thin films}},\ }\href {\doibase
  10.1103/PhysRevB.41.11203} {\bibfield  {journal} {\bibinfo  {journal} {Phys.
  Rev. B}\ }\textbf {\bibinfo {volume} {41}},\ \bibinfo {pages} {11203}
  (\bibinfo {year} {1990})}\BibitemShut {NoStop}%
\bibitem [{\citenamefont {Benkraouda}\ and\ \citenamefont
  {Clem}(1998)}]{Benk:1998}%
  \BibitemOpen
  \bibfield  {author} {\bibinfo {author} {\bibfnamefont {M.}~\bibnamefont
  {Benkraouda}}\ and\ \bibinfo {author} {\bibfnamefont {J.~R.}\ \bibnamefont
  {Clem}},\ }\bibfield  {title} {\emph {\bibinfo {title} {Critical current from
  surface barriers in type-{II} superconducting strips}},\ }\href {\doibase
  10.1103/PhysRevB.58.15103} {\bibfield  {journal} {\bibinfo  {journal} {Phys.
  Rev. B}\ }\textbf {\bibinfo {volume} {58}},\ \bibinfo {pages} {15103}
  (\bibinfo {year} {1998})}\BibitemShut {NoStop}%
\bibitem [{\citenamefont {Bean}(1962)}]{Bean:1962}%
  \BibitemOpen
  \bibfield  {author} {\bibinfo {author} {\bibfnamefont {C.~P.}\ \bibnamefont
  {Bean}},\ }\bibfield  {title} {\emph {\bibinfo {title} {Magnetization of hard
  superconductors}},\ }\href {\doibase 10.1103/PhysRevLett.8.250} {\bibfield
  {journal} {\bibinfo  {journal} {Phys. Rev. Lett.}\ }\textbf {\bibinfo
  {volume} {8}},\ \bibinfo {pages} {250} (\bibinfo {year} {1962})}\BibitemShut
  {NoStop}%
\bibitem [{\citenamefont {Bean}(1964)}]{Bean:1964}%
  \BibitemOpen
  \bibfield  {author} {\bibinfo {author} {\bibfnamefont {C.~P.}\ \bibnamefont
  {Bean}},\ }\bibfield  {title} {\emph {\bibinfo {title} {Magnetization of
  high-field superconductors}},\ }\href {\doibase 10.1103/RevModPhys.36.31}
  {\bibfield  {journal} {\bibinfo  {journal} {Rev. Mod. Phys.}\ }\textbf
  {\bibinfo {volume} {36}},\ \bibinfo {pages} {31} (\bibinfo {year}
  {1964})}\BibitemShut {NoStop}%
\bibitem [{\citenamefont {Schuster}\ \emph {et~al.}(1994)\citenamefont
  {Schuster}, \citenamefont {Indenbom}, \citenamefont {Kuhn}, \citenamefont
  {Brandt},\ and\ \citenamefont {Konczykowski}}]{Schuster:1994}%
  \BibitemOpen
  \bibfield  {author} {\bibinfo {author} {\bibfnamefont {T.}~\bibnamefont
  {Schuster}}, \bibinfo {author} {\bibfnamefont {M.~V.}\ \bibnamefont
  {Indenbom}}, \bibinfo {author} {\bibfnamefont {H.}~\bibnamefont {Kuhn}},
  \bibinfo {author} {\bibfnamefont {E.~H.}\ \bibnamefont {Brandt}}, \ and\
  \bibinfo {author} {\bibfnamefont {M.}~\bibnamefont {Konczykowski}},\
  }\bibfield  {title} {\emph {\bibinfo {title} {Flux penetration and
  overcritical currents in flat superconductors with irradiation-enhanced edge
  pinning: Theory and experiment}},\ }\href {\doibase
  10.1103/PhysRevLett.73.1424} {\bibfield  {journal} {\bibinfo  {journal}
  {Phys. Rev. Lett.}\ }\textbf {\bibinfo {volume} {73}},\ \bibinfo {pages}
  {1424} (\bibinfo {year} {1994})}\BibitemShut {NoStop}%
\bibitem [{\citenamefont {Ivanchenko}\ and\ \citenamefont
  {Mikheenko}(1983)}]{Ivanchenko:1983}%
  \BibitemOpen
  \bibfield  {author} {\bibinfo {author} {\bibfnamefont {Y.~M.}\ \bibnamefont
  {Ivanchenko}}\ and\ \bibinfo {author} {\bibfnamefont {P.~N.}\ \bibnamefont
  {Mikheenko}},\ }\bibfield  {title} {\emph {\bibinfo {title} {New mechanism of
  penetration of vortices into current-saturated superconducting films}},\
  }\href {http://www.jetp.ac.ru/cgi-bin/e/index/e/58/6/p1228?a=list} {\bibfield
   {journal} {\bibinfo  {journal} {Sov. Phys. JETP}\ }\textbf {\bibinfo
  {volume} {58}},\ \bibinfo {pages} {1228} (\bibinfo {year}
  {1983})}\BibitemShut {NoStop}%
\bibitem [{\citenamefont {Koshelev}\ and\ \citenamefont
  {Vinokur}(2001)}]{Koshelev:2001}%
  \BibitemOpen
  \bibfield  {author} {\bibinfo {author} {\bibfnamefont {A.~E.}\ \bibnamefont
  {Koshelev}}\ and\ \bibinfo {author} {\bibfnamefont {V.~M.}\ \bibnamefont
  {Vinokur}},\ }\bibfield  {title} {\emph {\bibinfo {title} {Suppression of
  surface barriers in superconductors by columnar defects}},\ }\href {\doibase
  10.1103/PhysRevB.64.134518} {\bibfield  {journal} {\bibinfo  {journal} {Phys.
  Rev. B}\ }\textbf {\bibinfo {volume} {64}},\ \bibinfo {pages} {134518}
  (\bibinfo {year} {2001})}\BibitemShut {NoStop}%
\bibitem [{\citenamefont {Gregory}\ \emph {et~al.}(2001)\citenamefont
  {Gregory}, \citenamefont {James}, \citenamefont {Bending}, \citenamefont
  {van~der Beek},\ and\ \citenamefont {Konczykowski}}]{Gregory:2001}%
  \BibitemOpen
  \bibfield  {author} {\bibinfo {author} {\bibfnamefont {J.~K.}\ \bibnamefont
  {Gregory}}, \bibinfo {author} {\bibfnamefont {M.~S.}\ \bibnamefont {James}},
  \bibinfo {author} {\bibfnamefont {S.~J.}\ \bibnamefont {Bending}}, \bibinfo
  {author} {\bibfnamefont {C.~J.}\ \bibnamefont {van~der Beek}}, \ and\
  \bibinfo {author} {\bibfnamefont {M.}~\bibnamefont {Konczykowski}},\
  }\bibfield  {title} {\emph {\bibinfo {title} {Suppression of surface barriers
  for flux penetration in {Bi$_2$Sr$_2$CaCu$_2$O$_{8+\delta}$} whiskers by
  electron and heavy ion irradiation}},\ }\href {\doibase
  10.1103/PhysRevB.64.134517} {\bibfield  {journal} {\bibinfo  {journal} {Phys.
  Rev. B}\ }\textbf {\bibinfo {volume} {64}},\ \bibinfo {pages} {134517}
  (\bibinfo {year} {2001})}\BibitemShut {NoStop}%
\bibitem [{\citenamefont {Bush}\ \emph {et~al.}(2002)\citenamefont {Bush},
  \citenamefont {Dorofeev}, \citenamefont {Drobin}, \citenamefont {Kuroedov},
  \citenamefont {Vladumirova},\ and\ \citenamefont {Vyatkin}}]{Bush:2002}%
  \BibitemOpen
  \bibfield  {author} {\bibinfo {author} {\bibfnamefont {A.~A.}\ \bibnamefont
  {Bush}}, \bibinfo {author} {\bibfnamefont {G.~L.}\ \bibnamefont {Dorofeev}},
  \bibinfo {author} {\bibfnamefont {V.~M.}\ \bibnamefont {Drobin}}, \bibinfo
  {author} {\bibfnamefont {Y.~D.}\ \bibnamefont {Kuroedov}}, \bibinfo {author}
  {\bibfnamefont {N.~M.}\ \bibnamefont {Vladumirova}}, \ and\ \bibinfo {author}
  {\bibfnamefont {V.~S.}\ \bibnamefont {Vyatkin}},\ }\bibfield  {title} {\emph
  {\bibinfo {title} {{Bean-Livingston} barrier and dynamics of the magnetic
  flux flow in layered (plated) superconductors}},\ }\href {\doibase
  10.1109/TASC.2002.1018573} {\bibfield  {journal} {\bibinfo  {journal} {IEEE
  Trans. Appl. Supercond.}\ }\textbf {\bibinfo {volume} {12}},\ \bibinfo
  {pages} {1018} (\bibinfo {year} {2002})}\BibitemShut {NoStop}%
\bibitem [{\citenamefont {Ovchinnikov}\ and\ \citenamefont
  {Varlamov}(2015)}]{Ovchinnikov:2015}%
  \BibitemOpen
  \bibfield  {author} {\bibinfo {author} {\bibfnamefont {Y.~N.}\ \bibnamefont
  {Ovchinnikov}}\ and\ \bibinfo {author} {\bibfnamefont {A.~A.}\ \bibnamefont
  {Varlamov}},\ }\bibfield  {title} {\emph {\bibinfo {title} {Phase slips in a
  current-biased narrow superconducting strip}},\ }\href {\doibase
  10.1103/PhysRevB.91.014514} {\bibfield  {journal} {\bibinfo  {journal} {Phys.
  Rev. B}\ }\textbf {\bibinfo {volume} {91}},\ \bibinfo {pages} {014514}
  (\bibinfo {year} {2015})}\BibitemShut {NoStop}%
\bibitem [{\citenamefont {Kimmel}\ \emph
  {et~al.}(2017{\natexlab{a}})\citenamefont {Kimmel}, \citenamefont {Glatz},\
  and\ \citenamefont {Aranson}}]{Kimmel:2017b}%
  \BibitemOpen
  \bibfield  {author} {\bibinfo {author} {\bibfnamefont {G.}~\bibnamefont
  {Kimmel}}, \bibinfo {author} {\bibfnamefont {A.}~\bibnamefont {Glatz}}, \
  and\ \bibinfo {author} {\bibfnamefont {I.~S.}\ \bibnamefont {Aranson}},\
  }\bibfield  {title} {\emph {\bibinfo {title} {Phase slips in superconducting
  weak links}},\ }\href {\doibase 10.1103/PhysRevB.95.014518} {\bibfield
  {journal} {\bibinfo  {journal} {Phys. Rev. B}\ }\textbf {\bibinfo {volume}
  {95}},\ \bibinfo {pages} {014518} (\bibinfo {year}
  {2017}{\natexlab{a}})}\BibitemShut {NoStop}%
\bibitem [{\citenamefont {Sadovskyy}\ \emph {et~al.}(2015)\citenamefont
  {Sadovskyy}, \citenamefont {Koshelev}, \citenamefont {Phillips},
  \citenamefont {Karpeyev},\ and\ \citenamefont {Glatz}}]{Sadovskyy:2015a}%
  \BibitemOpen
  \bibfield  {author} {\bibinfo {author} {\bibfnamefont {I.~A.}\ \bibnamefont
  {Sadovskyy}}, \bibinfo {author} {\bibfnamefont {A.~E.}\ \bibnamefont
  {Koshelev}}, \bibinfo {author} {\bibfnamefont {C.~L.}\ \bibnamefont
  {Phillips}}, \bibinfo {author} {\bibfnamefont {D.~A.}\ \bibnamefont
  {Karpeyev}}, \ and\ \bibinfo {author} {\bibfnamefont {A.}~\bibnamefont
  {Glatz}},\ }\bibfield  {title} {\emph {\bibinfo {title} {Stable large-scale
  solver for {Ginzburg-Landau} equations for superconductors}},\ }\href
  {\doibase 10.1016/j.jcp.2015.04.002} {\bibfield  {journal} {\bibinfo
  {journal} {J. Comp. Phys.}\ }\textbf {\bibinfo {volume} {294}},\ \bibinfo
  {pages} {639} (\bibinfo {year} {2015})}\BibitemShut {NoStop}%
\bibitem [{\citenamefont {Berdiyorov}\ \emph {et~al.}(2006)\citenamefont
  {Berdiyorov}, \citenamefont {Milo\v{s}evi\'{c}},\ and\ \citenamefont
  {Peeters}}]{Berdiyorov:2006}%
  \BibitemOpen
  \bibfield  {author} {\bibinfo {author} {\bibfnamefont {G.~R.}\ \bibnamefont
  {Berdiyorov}}, \bibinfo {author} {\bibfnamefont {M.~V.}\ \bibnamefont
  {Milo\v{s}evi\'{c}}}, \ and\ \bibinfo {author} {\bibfnamefont {F.~M.}\
  \bibnamefont {Peeters}},\ }\bibfield  {title} {\emph {\bibinfo {title} {Novel
  commensurability effects in superconducting films with antidot arrays}},\
  }\href {\doibase 10.1103/PhysRevLett.96.207001} {\bibfield  {journal}
  {\bibinfo  {journal} {Phys. Rev. Lett.}\ }\textbf {\bibinfo {volume} {96}},\
  \bibinfo {pages} {207001} (\bibinfo {year} {2006})}\BibitemShut {NoStop}%
\bibitem [{\citenamefont {Sadovskyy}\ \emph
  {et~al.}(2016{\natexlab{a}})\citenamefont {Sadovskyy}, \citenamefont
  {Koshelev}, \citenamefont {Glatz}, \citenamefont {Ortalan}, \citenamefont
  {Rupich},\ and\ \citenamefont {Leroux}}]{Sadovskyy:2016a}%
  \BibitemOpen
  \bibfield  {author} {\bibinfo {author} {\bibfnamefont {I.~A.}\ \bibnamefont
  {Sadovskyy}}, \bibinfo {author} {\bibfnamefont {A.~E.}\ \bibnamefont
  {Koshelev}}, \bibinfo {author} {\bibfnamefont {A.}~\bibnamefont {Glatz}},
  \bibinfo {author} {\bibfnamefont {V.}~\bibnamefont {Ortalan}}, \bibinfo
  {author} {\bibfnamefont {M.~W.}\ \bibnamefont {Rupich}}, \ and\ \bibinfo
  {author} {\bibfnamefont {M.}~\bibnamefont {Leroux}},\ }\bibfield  {title}
  {\emph {\bibinfo {title} {Simulation of the vortex dynamics in a real pinning
  landscape of {YBa$_2$Cu$_3$O$_{7-\delta}$} coated conductors}},\ }\href
  {\doibase 10.1103/PhysRevApplied.5.014011} {\bibfield  {journal} {\bibinfo
  {journal} {Phys. Rev. Applied}\ }\textbf {\bibinfo {volume} {5}},\ \bibinfo
  {pages} {014011} (\bibinfo {year} {2016}{\natexlab{a}})}\BibitemShut
  {NoStop}%
\bibitem [{\citenamefont {Sadovskyy}\ \emph
  {et~al.}(2016{\natexlab{b}})\citenamefont {Sadovskyy}, \citenamefont {Jia},
  \citenamefont {Leroux}, \citenamefont {Kwon}, \citenamefont {Hu},
  \citenamefont {Fang}, \citenamefont {Chaparro}, \citenamefont {Zhu},
  \citenamefont {Welp}, \citenamefont {Zuo}, \citenamefont {Zhang},
  \citenamefont {Nakasaki}, \citenamefont {Selvamanickam}, \citenamefont
  {Crabtree}, \citenamefont {Koshelev}, \citenamefont {Glatz},\ and\
  \citenamefont {Kwok}}]{Sadovskyy:2016b}%
  \BibitemOpen
  \bibfield  {author} {\bibinfo {author} {\bibfnamefont {I.~A.}\ \bibnamefont
  {Sadovskyy}}, \bibinfo {author} {\bibfnamefont {Y.}~\bibnamefont {Jia}},
  \bibinfo {author} {\bibfnamefont {M.}~\bibnamefont {Leroux}}, \bibinfo
  {author} {\bibfnamefont {J.}~\bibnamefont {Kwon}}, \bibinfo {author}
  {\bibfnamefont {H.}~\bibnamefont {Hu}}, \bibinfo {author} {\bibfnamefont
  {L.}~\bibnamefont {Fang}}, \bibinfo {author} {\bibfnamefont {C.}~\bibnamefont
  {Chaparro}}, \bibinfo {author} {\bibfnamefont {S.}~\bibnamefont {Zhu}},
  \bibinfo {author} {\bibfnamefont {U.}~\bibnamefont {Welp}}, \bibinfo {author}
  {\bibfnamefont {J.-M.}\ \bibnamefont {Zuo}}, \bibinfo {author} {\bibfnamefont
  {Y.}~\bibnamefont {Zhang}}, \bibinfo {author} {\bibfnamefont
  {R.}~\bibnamefont {Nakasaki}}, \bibinfo {author} {\bibfnamefont
  {V.}~\bibnamefont {Selvamanickam}}, \bibinfo {author} {\bibfnamefont {G.~W.}\
  \bibnamefont {Crabtree}}, \bibinfo {author} {\bibfnamefont {A.~E.}\
  \bibnamefont {Koshelev}}, \bibinfo {author} {\bibfnamefont {A.}~\bibnamefont
  {Glatz}}, \ and\ \bibinfo {author} {\bibfnamefont {W.-K.}\ \bibnamefont
  {Kwok}},\ }\bibfield  {title} {\emph {\bibinfo {title} {Toward
  superconducting critical current by design}},\ }\href {\doibase
  10.1002/adma.201600602} {\bibfield  {journal} {\bibinfo  {journal} {Adv.
  Mater.}\ }\textbf {\bibinfo {volume} {28}},\ \bibinfo {pages} {4593}
  (\bibinfo {year} {2016}{\natexlab{b}})}\BibitemShut {NoStop}%
\bibitem [{\citenamefont {Sadovskyy}\ \emph {et~al.}(2017)\citenamefont
  {Sadovskyy}, \citenamefont {Wang}, \citenamefont {Xiao}, \citenamefont
  {Kwok},\ and\ \citenamefont {Glatz}}]{Sadovskyy:2017}%
  \BibitemOpen
  \bibfield  {author} {\bibinfo {author} {\bibfnamefont {I.~A.}\ \bibnamefont
  {Sadovskyy}}, \bibinfo {author} {\bibfnamefont {Y.~L.}\ \bibnamefont {Wang}},
  \bibinfo {author} {\bibfnamefont {Z.-L.}\ \bibnamefont {Xiao}}, \bibinfo
  {author} {\bibfnamefont {W.-K.}\ \bibnamefont {Kwok}}, \ and\ \bibinfo
  {author} {\bibfnamefont {A.}~\bibnamefont {Glatz}},\ }\bibfield  {title}
  {\emph {\bibinfo {title} {Effect of hexagonal patterned arrays and defect
  geometry on the critical current of superconducting films}},\ }\href
  {\doibase 10.1103/PhysRevB.95.075303} {\bibfield  {journal} {\bibinfo
  {journal} {Phys. Rev. B}\ }\textbf {\bibinfo {volume} {95}},\ \bibinfo
  {pages} {075303} (\bibinfo {year} {2017})}\BibitemShut {NoStop}%
\bibitem [{\citenamefont {Brandt}\ and\ \citenamefont
  {Indenbom}(1993)}]{Brandt:1993}%
  \BibitemOpen
  \bibfield  {author} {\bibinfo {author} {\bibfnamefont {E.~H.}\ \bibnamefont
  {Brandt}}\ and\ \bibinfo {author} {\bibfnamefont {M.}~\bibnamefont
  {Indenbom}},\ }\bibfield  {title} {\emph {\bibinfo {title}
  {Type-{II}-superconductor strip with current in a perpendicular magnetic
  field}},\ }\href {\doibase 10.1103/PhysRevB.48.12893} {\bibfield  {journal}
  {\bibinfo  {journal} {Phys. Rev. B}\ }\textbf {\bibinfo {volume} {48}},\
  \bibinfo {pages} {12893} (\bibinfo {year} {1993})}\BibitemShut {NoStop}%
\bibitem [{\citenamefont {Burlachkov}\ \emph {et~al.}(1991)\citenamefont
  {Burlachkov}, \citenamefont {Konczykowski}, \citenamefont {Yeshurun},\ and\
  \citenamefont {Holtzberg}}]{Burlachkov:1991}%
  \BibitemOpen
  \bibfield  {author} {\bibinfo {author} {\bibfnamefont {L.}~\bibnamefont
  {Burlachkov}}, \bibinfo {author} {\bibfnamefont {M.}~\bibnamefont
  {Konczykowski}}, \bibinfo {author} {\bibfnamefont {Y.}~\bibnamefont
  {Yeshurun}}, \ and\ \bibinfo {author} {\bibfnamefont {F.}~\bibnamefont
  {Holtzberg}},\ }\bibfield  {title} {\emph {\bibinfo {title}
  {{Bean-Livingston} barriers and first field for flux penetration in
  high-{$T_\mathrm{c}$} crystals}},\ }\href {\doibase 10.1063/1.350152}
  {\bibfield  {journal} {\bibinfo  {journal} {J. Appl. Phys.}\ }\textbf
  {\bibinfo {volume} {70}},\ \bibinfo {pages} {5759} (\bibinfo {year}
  {1991})}\BibitemShut {NoStop}%
\bibitem [{\citenamefont {Burlachkov}(1993)}]{Burlachkov:1993}%
  \BibitemOpen
  \bibfield  {author} {\bibinfo {author} {\bibfnamefont {L.}~\bibnamefont
  {Burlachkov}},\ }\bibfield  {title} {\emph {\bibinfo {title} {Magnetic
  relaxation over the {Bean-Livingston} surface barrier}},\ }\href {\doibase
  10.1103/PhysRevB.47.8056} {\bibfield  {journal} {\bibinfo  {journal} {Phys.
  Rev. B}\ }\textbf {\bibinfo {volume} {47}},\ \bibinfo {pages} {8056}
  (\bibinfo {year} {1993})}\BibitemShut {NoStop}%
\bibitem [{\citenamefont {Burlachkov}\ \emph {et~al.}(1996)\citenamefont
  {Burlachkov}, \citenamefont {Koshelev},\ and\ \citenamefont
  {Vinokur}}]{Burlachkov:1996}%
  \BibitemOpen
  \bibfield  {author} {\bibinfo {author} {\bibfnamefont {L.}~\bibnamefont
  {Burlachkov}}, \bibinfo {author} {\bibfnamefont {A.~E.}\ \bibnamefont
  {Koshelev}}, \ and\ \bibinfo {author} {\bibfnamefont {V.~M.}\ \bibnamefont
  {Vinokur}},\ }\bibfield  {title} {\emph {\bibinfo {title} {Transport
  properties of high-temperature superconductors: Surface vs bulk effect}},\
  }\href {\doibase 10.1103/PhysRevB.54.6750} {\bibfield  {journal} {\bibinfo
  {journal} {Phys. Rev. B}\ }\textbf {\bibinfo {volume} {54}},\ \bibinfo
  {pages} {6750} (\bibinfo {year} {1996})}\BibitemShut {NoStop}%
\bibitem [{\citenamefont {Koshelev}\ \emph {et~al.}(2016)\citenamefont
  {Koshelev}, \citenamefont {Sadovskyy}, \citenamefont {Phillips},\ and\
  \citenamefont {Glatz}}]{Koshelev:2016}%
  \BibitemOpen
  \bibfield  {author} {\bibinfo {author} {\bibfnamefont {A.~E.}\ \bibnamefont
  {Koshelev}}, \bibinfo {author} {\bibfnamefont {I.~A.}\ \bibnamefont
  {Sadovskyy}}, \bibinfo {author} {\bibfnamefont {C.~L.}\ \bibnamefont
  {Phillips}}, \ and\ \bibinfo {author} {\bibfnamefont {A.}~\bibnamefont
  {Glatz}},\ }\bibfield  {title} {\emph {\bibinfo {title} {Optimization of
  vortex pinning by nanoparticles using simulations of the time-dependent
  {Ginzburg-Landau} model}},\ }\href {\doibase 10.1103/PhysRevB.93.060508}
  {\bibfield  {journal} {\bibinfo  {journal} {Phys. Rev. B}\ }\textbf {\bibinfo
  {volume} {93}},\ \bibinfo {pages} {060508} (\bibinfo {year}
  {2016})}\BibitemShut {NoStop}%
\bibitem [{\citenamefont {Willa}\ \emph
  {et~al.}(2018{\natexlab{a}})\citenamefont {Willa}, \citenamefont {Koshelev},
  \citenamefont {Sadovskyy},\ and\ \citenamefont {Glatz}}]{Willa:2018a}%
  \BibitemOpen
  \bibfield  {author} {\bibinfo {author} {\bibfnamefont {R.}~\bibnamefont
  {Willa}}, \bibinfo {author} {\bibfnamefont {A.~E.}\ \bibnamefont {Koshelev}},
  \bibinfo {author} {\bibfnamefont {I.~A.}\ \bibnamefont {Sadovskyy}}, \ and\
  \bibinfo {author} {\bibfnamefont {A.}~\bibnamefont {Glatz}},\ }\bibfield
  {title} {\emph {\bibinfo {title} {Strong-pinning regimes by spherical
  inclusions in anisotropic {type-II} superconductors}},\ }\href {\doibase
  10.1088/1361-6668/aa939e} {\bibfield  {journal} {\bibinfo  {journal}
  {Supercond. Sci. Technol.}\ }\textbf {\bibinfo {volume} {31}},\ \bibinfo
  {pages} {014001} (\bibinfo {year} {2018}{\natexlab{a}})}\BibitemShut
  {NoStop}%
\bibitem [{\citenamefont {Willa}\ \emph
  {et~al.}(2018{\natexlab{b}})\citenamefont {Willa}, \citenamefont {Koshelev},
  \citenamefont {Sadovskyy},\ and\ \citenamefont {Glatz}}]{Willa:2018b}%
  \BibitemOpen
  \bibfield  {author} {\bibinfo {author} {\bibfnamefont {R.}~\bibnamefont
  {Willa}}, \bibinfo {author} {\bibfnamefont {A.~E.}\ \bibnamefont {Koshelev}},
  \bibinfo {author} {\bibfnamefont {I.~A.}\ \bibnamefont {Sadovskyy}}, \ and\
  \bibinfo {author} {\bibfnamefont {A.}~\bibnamefont {Glatz}},\ }\bibfield
  {title} {\emph {\bibinfo {title} {Peak effect due to competing vortex ground
  states in superconductors with large inclusions}},\ }\href {\doibase
  10.1103/PhysRevB.98.054517} {\bibfield  {journal} {\bibinfo  {journal} {Phys.
  Rev. B}\ }\textbf {\bibinfo {volume} {98}},\ \bibinfo {pages} {054517}
  (\bibinfo {year} {2018}{\natexlab{b}})}\BibitemShut {NoStop}%
\bibitem [{\citenamefont {Kimmel}\ \emph
  {et~al.}(2017{\natexlab{b}})\citenamefont {Kimmel}, \citenamefont
  {Sadovskyy},\ and\ \citenamefont {Glatz}}]{Kimmel:2017a}%
  \BibitemOpen
  \bibfield  {author} {\bibinfo {author} {\bibfnamefont {G.}~\bibnamefont
  {Kimmel}}, \bibinfo {author} {\bibfnamefont {I.~A.}\ \bibnamefont
  {Sadovskyy}}, \ and\ \bibinfo {author} {\bibfnamefont {A.}~\bibnamefont
  {Glatz}},\ }\bibfield  {title} {\emph {\bibinfo {title} {In silico
  optimization of critical currents in superconductors}},\ }\href {\doibase
  10.1103/PhysRevE.96.013318} {\bibfield  {journal} {\bibinfo  {journal} {Phys.
  Rev. E}\ }\textbf {\bibinfo {volume} {96}},\ \bibinfo {pages} {013318}
  (\bibinfo {year} {2017}{\natexlab{b}})}\BibitemShut {NoStop}%
\bibitem [{\citenamefont {Besseling}\ \emph {et~al.}(2005)\citenamefont
  {Besseling}, \citenamefont {Kes}, \citenamefont {Dr\"{o}se},\ and\
  \citenamefont {Vinokur}}]{Besseling:2005}%
  \BibitemOpen
  \bibfield  {author} {\bibinfo {author} {\bibfnamefont {R.}~\bibnamefont
  {Besseling}}, \bibinfo {author} {\bibfnamefont {P.~H.}\ \bibnamefont {Kes}},
  \bibinfo {author} {\bibfnamefont {T.}~\bibnamefont {Dr\"{o}se}}, \ and\
  \bibinfo {author} {\bibfnamefont {V.~M.}\ \bibnamefont {Vinokur}},\
  }\bibfield  {title} {\emph {\bibinfo {title} {Depinning and dynamics of
  vortices confined in mesoscopic flow channels}},\ }\href {\doibase
  10.1088/1367-2630/7/1/071} {\bibfield  {journal} {\bibinfo  {journal} {New J.
  Phys.}\ }\textbf {\bibinfo {volume} {7}},\ \bibinfo {pages} {71} (\bibinfo
  {year} {2005})}\BibitemShut {NoStop}%
\end{thebibliography}%

\end{document}